\documentclass[preprint]{aastex631}

\newcommand{\objname}{362P}
\newcommand{\objnameFull}{362P/(457175) 2008 GO$_{98}$}
\usepackage{placeins}
\usepackage{soul}
\usepackage[outline]{contour}
\usepackage[percent]{overpic}
\newcommand{\labelcolor}{white} 

\newcommand{\labelpicA}[5]{
\begin{overpic}[width=#4\linewidth]{#1}
	\put (5,7) {\huge\color{\labelcolor} \textbf{\contour{black}{#2}}}
	\put (45,8) {\large\color{\labelcolor} \textbf{\contour{black}{#3}}}
	\put (2,52) {\includegraphics[width=0.11\linewidth]{#5}}
\end{overpic}
}

\begin{document}

\title{Recurrent Cometary Activity Discovered on Quasi-Hilda Jupiter Family Comet\\ 362P/(457175) 2008 GO$_{98}$}

\correspondingauthor{Kennedy A. Farrell}
\email{kaf435@nau.edu}

\author[0000-0003-2521-848X]{Kennedy A. Farrell}
\affiliation{Dept. of Astronomy \& Planetary Science, Northern Arizona University, PO Box 6010, Flagstaff, AZ 86011, USA}
\affiliation{Lowell Observatory, 1400 W Mars Hill Road, Flagstaff, AZ 86001, USA}

\author[0000-0001-7335-1715]{Colin Orion Chandler}
\affiliation{Dept. of Astronomy \& the DiRAC Institute, University of Washington, 3910 15th Ave NE, Seattle, WA 98195, USA}
\affiliation{LSST Interdisciplinary Network for Collaboration and Computing, 933 N. Cherry Avenue, Tucson, AZ 85721, USA}
\affiliation{Dept. of Astronomy \& Planetary Science, Northern Arizona University, PO Box 6010, Flagstaff, AZ 86011, USA}

\author[0000-0001-9859-0894]{Chadwick A. Trujillo}
\affiliation{Dept. of Astronomy \& Planetary Science, Northern Arizona University, PO Box 6010, Flagstaff, AZ 86011, USA}

\author[0000-0001-5750-4953]{William J. Oldroyd}
\affiliation{Dept. of Astronomy \& Planetary Science, Northern Arizona University, PO Box 6010, Flagstaff, AZ 86011, USA}

\author[0000-0002-7489-5893]{Jarod A. DeSpain}
\affiliation{Dept. of Astronomy \& Planetary Science, Northern Arizona University, PO Box 6010, Flagstaff, AZ 86011, USA}

\author[0000-0003-2113-3593]{Mark Jesus Mendoza Magbanua}
\affiliation{Dept. of Laboratory Medicine, University of California San Francisco, 2340 Sutter Street, San Francisco, CA 94143, USA}

\author[0000-0002-8069-3139]{Maxwell K. Frissell}
\affiliation{Dept. of Astronomy \& the DiRAC Institute, University of Washington, 3910 15th Ave NE, Seattle, WA 98195, USA}
\affiliation{Dept. of Astronomy \& Planetary Science, Northern Arizona University, PO Box 6010, Flagstaff, AZ 86011, USA}

\author[0009-0003-1154-6578]{Phineas Stone} 
\affiliation{Dept. of Astronomy \& the DiRAC Institute, University of Washington, 3910 15th Ave NE, Seattle, WA 98195, USA}

\begin{abstract}
We report the discovery of recurrent activity on quasi-Hilda comet (QHC) \objnameFull{}. The first activity epoch was discovered during the perihelion passage of \objname{} in 2016 \citep{GARCIAMIGANI201812}, so we were motivated to observe it for recurrent cometary activity near its next perihelion passage (UT 2024 July 20). We obtained observations with the Lowell Discovery Telescope (LDT), the Astrophysical Research Consortium (ARC) telescope, and the Vatican Advanced Technology Telescope (VATT) and identified a second activity epoch when \objname{} had a true anomaly ($\nu$) as early as 318.1$^\circ$. We conducted archival searches of numerous repositories and identified images obtained with Canada-France-Hawaii Telescope (CFHT) MegaCam, Dark Energy Camera (DECam), Pan-STARRS 1, SkyMapper, Zwicky Transient Facility (ZTF), and Las Cumbres Observatory Global Telescope (LCOGT) network data. Using these data, we identified activity from a previously unreported timespan, and we did not detect activity when \objname{} was away from perihelion, specifically 83$^\circ$$<\nu<$318$^\circ$. Detection of activity near perihelion and absence of activity away from perihelion suggest thermally-driven activity and volatile sublimation. Our dynamical simulations suggest \objname{} is a QHC and it will remain in a combined Jupiter-family comet (JFC) and quasi-Hilda orbit over the next 1~kyr, though it will become increasingly chaotic nearing the end of this timeframe. Our backward simulations suggest \objname{} may have migrated from the orbit of a Long Period Comet ($\sim$53\%) or Centaur ($\sim$32\%), otherwise it remained a JFC ($\sim$15\%) over the previous 100~kyr. We recommend additional telescope observations from the community as \objname{} continues outbound from its perihelion on UT 2024 July 20, as well as continued observations for a third activity epoch.
\end{abstract}


\section{Introduction}\label{sec:intro}
The term Jupiter-family comet (JFC) describes a dynamical class of active objects in our solar system. JFCs have periods of $<20$~yr \citep{1996ASPC..107..173L} and appear with cometary activity in the form of a tail or coma. Activity from JFCs suggests volatile sublimation during periods of increased solar heating \citep{jewitt2022asteroidcomet} when closer in proximity to the Sun. A JFC experiencing volatile sublimation may increase in brightness when solar heating is near maximum, which occurs during the perihelion passage of the comet.

Gravitational interactions with giant planets are responsible for the dynamical migration of JFCs from orbits beyond Neptune to orbits near Jupiter \citep{fraser2022transition}. Continued gravitational interactions with Jupiter can transition JFCs into the dynamically stable 3:2 mean-motion resonance of the Hilda group \citep{2005Icar..174...81D}. The Hildas are a dynamically similar population of asteroids, referred to as Hilda asteroids or Hilda objects. Alternatively, the quasi-Hildas are a non-exclusive population of asteroids and comets that do not fall into the 3:2 mean-motion resonance with Jupiter. JFCs that fall near, but not into, this stable resonance may orbit with the dynamically unstable quasi-Hilda group as quasi-Hilda Comets (QHCs; \citealt{2016A&A...590A.111G, 2006A&A...448.1191T}).

Quasi-Hildas offer an opportunity to learn more about active objects and their dynamical interactions in the solar system, given the relatively high activity rate among quasi-Hildas \citep[15:300][]{chandlerMigratoryOutburstingQuasiHilda2022, 2016A&A...590A.111G, gil-hutton2023} compared to the activity rate found among Main-Belt Asteroids (MBAs) \citep[1:10,000][]{jewittActiveAsteroids2015, hsiehMainbeltCometsPanSTARRS12015, Chandler_2024}. Because of their dynamical instability, quasi-Hildas can transition to other dynamical classes after gravitational perturbations from Jupiter \citep{chandlerMigratoryOutburstingQuasiHilda2022, 2005Icar..174...81D, 2019Icar..319..828D}.

\objnameFull{}, hereafter \objname{}, was first discovered on UT 2008 April 8 ($r=$ 3.246~au, $\nu=$ 332.3$^\circ$) by Spacewatch with the 0.9~m telescope at Kitt Peak National Observatory (Tucson, Arizona). Observations for the previously discovered epoch of activity discussed in \cite{GARCIAMIGANI201812} began after \objname{} had passed perihelion on UT 2016 August 23 \citep[$r=$ 2.850~au, $\nu=$ 360.0$^\circ$;][]{giorginiJPLOnLineSolar1996}. Activity originating from the nucleus was discovered as a coma, though activity evolved into a tail as observations continued past the 2016 perihelion passage of \objname{} and into the outgoing perihelion arc before concluding near UT 2017 July 25 \citep[$r=$ 3.326~au and $\nu=$ 69.69$^\circ$;][]{GARCIAMIGANI201812}. Activity indicators from \objname{} are significant to studies of the population of known active objects in the solar system \citep{hsiehMainbeltCometsPanSTARRS12015, chandlerMigratoryOutburstingQuasiHilda2022, Chandler_2024}. After cometary activity was discovered from \objname{}, it was designated as a JFC with a quasi-Hilda orbit \citep{GARCIAMIGANI201812}.

A second activity epoch had not been observed for \objname{}, so we conducted additional telescope observations, discussed in Section \ref{sec:telobs}. To support our understanding of the activity pattern of \objname{}, we also explored archival data, the thermodynamics of the comet, and its observability. To further explore the dynamical journey of \objname{} to a quasi-Hilda orbit and future dynamical evolution, we carried out dynamical simulations detailed in Section \ref{sec:dynamics}.

\section{Observations}\label{sec:telobs}
Our detection of recurrent cometary sublimation near perihelion from \objname{} is bolstered by our follow-up telescope observations (Table \ref{tab:obs}) and an archival search. We find evidence of an activity pattern in images from our observations, archival images from the first activity epoch, and archival images since the first epoch in addition to that reported in \cite{GARCIAMIGANI201812} and \cite{gil-hutton2023}. We note and discuss an apparent evolution of activity over the outgoing perihelion arc, as \objname{} appeared with a linear tail that became more diffuse.

\newcommand{\figsize}{0.3}
\begin{figure*}[ht]
\centering
\label{fig:obs}
\begin{tabular}{ccc}
        \vspace{2pt}
        \labelpicA{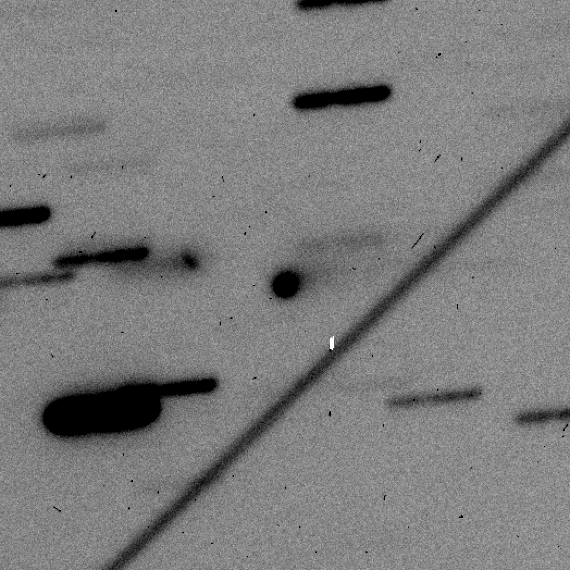}{a}{2024-01-12}{\figsize}{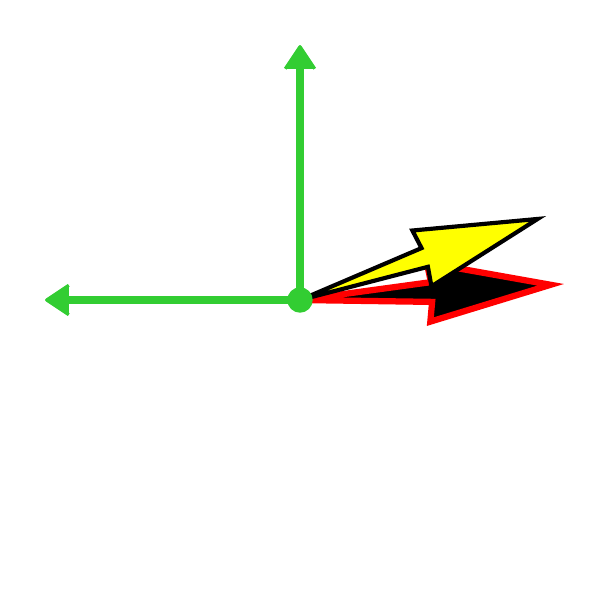} & 
        \labelpicA{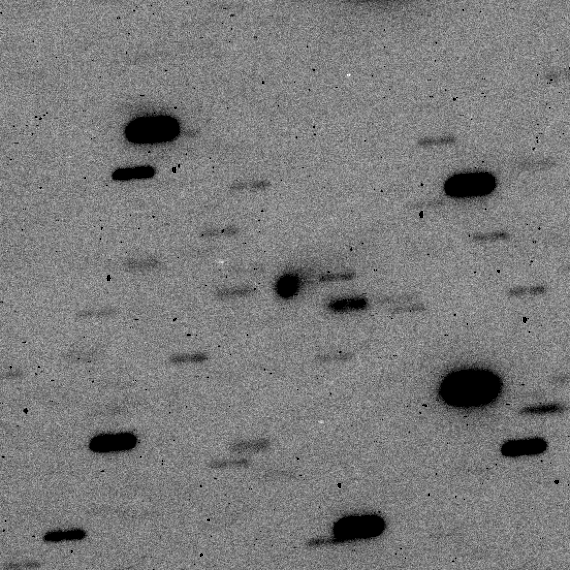}{b}{2024-01-13}{\figsize}{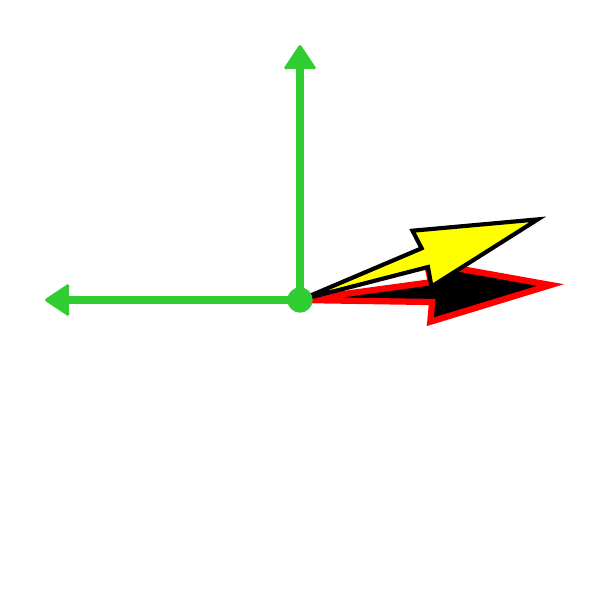} & 
        \labelpicA{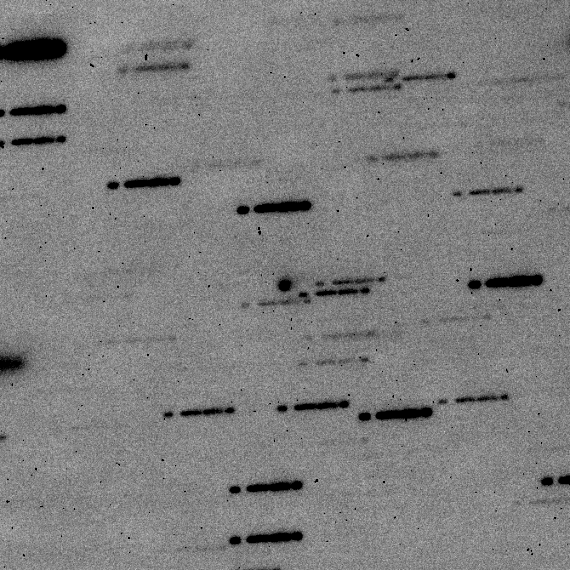}{c}{2024-01-16}{\figsize}{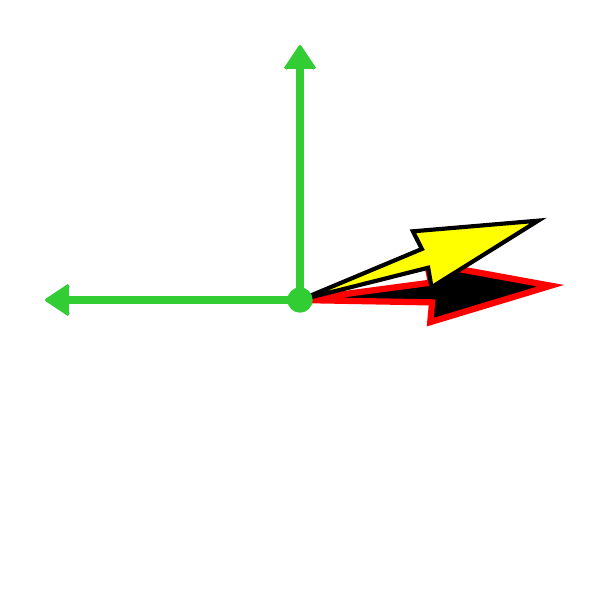} 
        \\
        \labelpicA{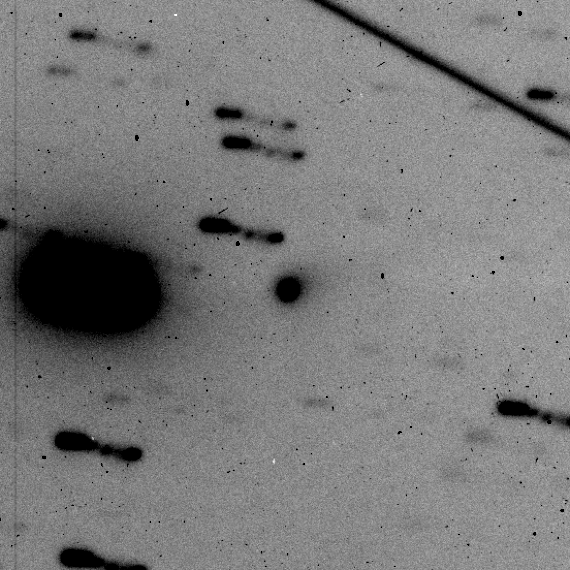}{d}{2024-02-15}{\figsize}{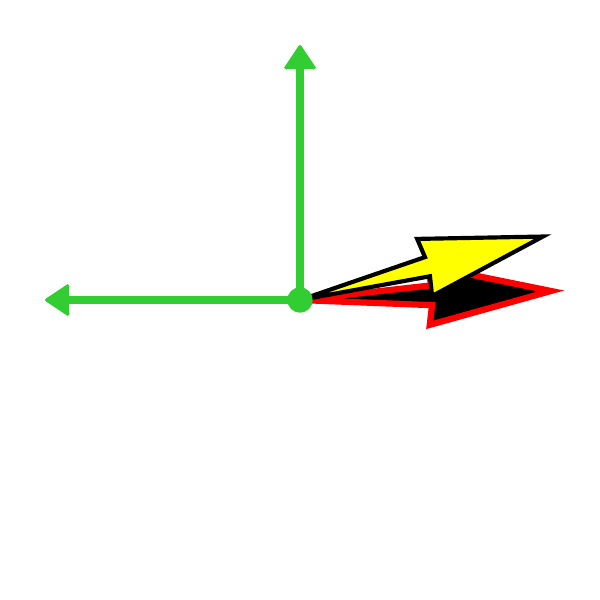} &
        \labelpicA{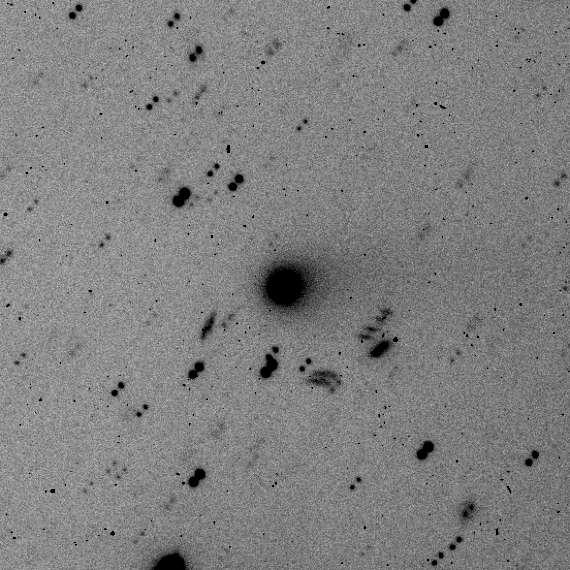}{e}{2024-04-11}{\figsize}{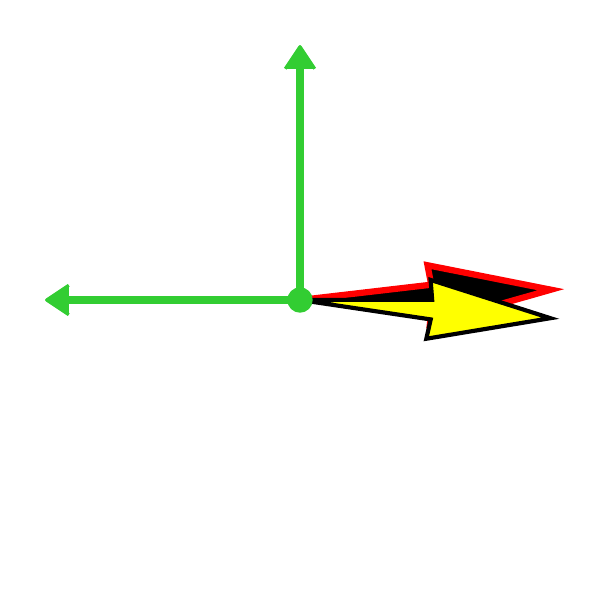} &
        \labelpicA{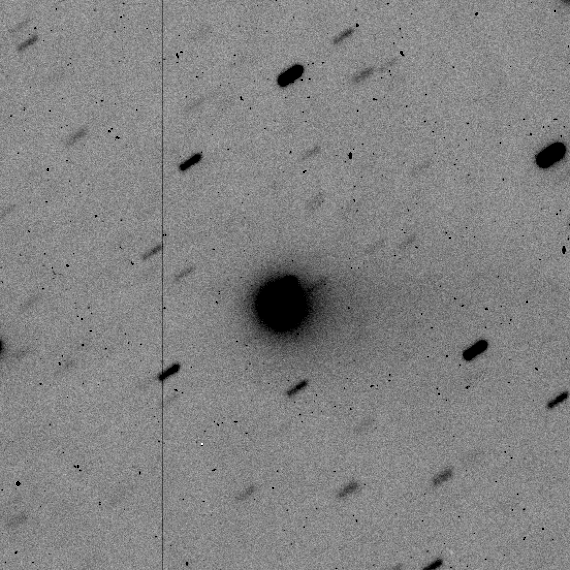}{f}{2024-05-09}{\figsize}{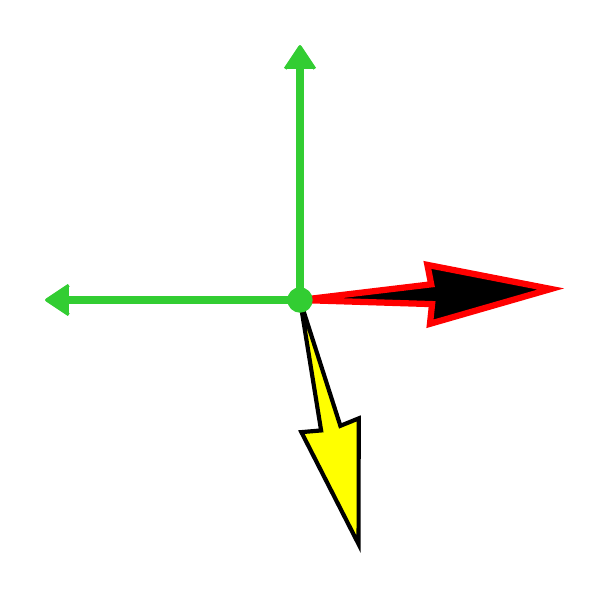}
        \\
        \labelpicA{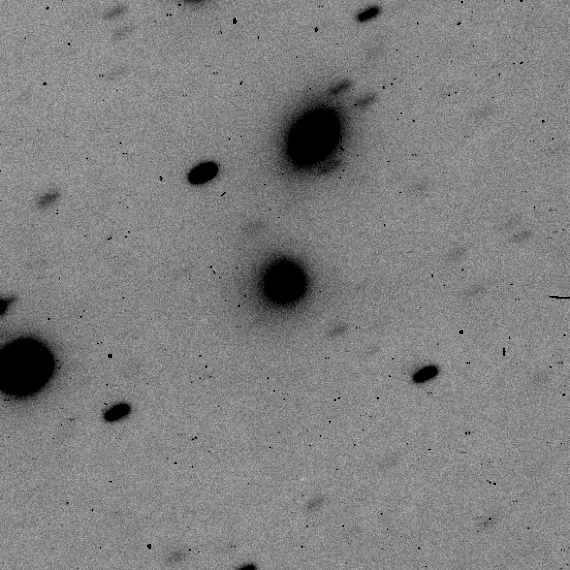}{g}{2024-08-09}{\figsize}{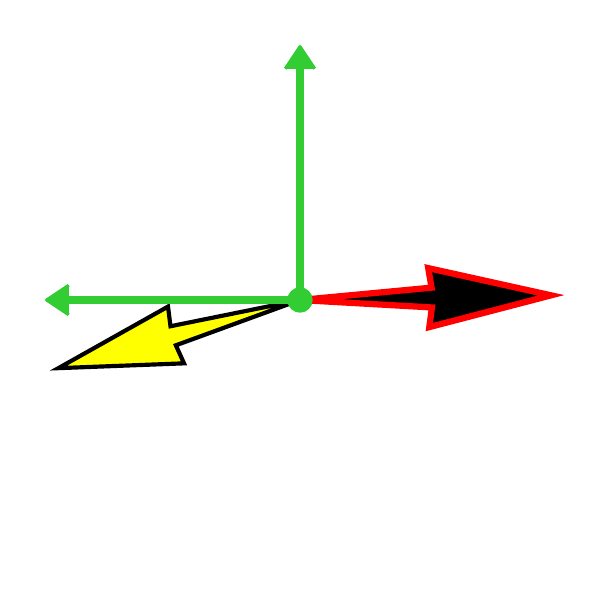} &
        \labelpicA{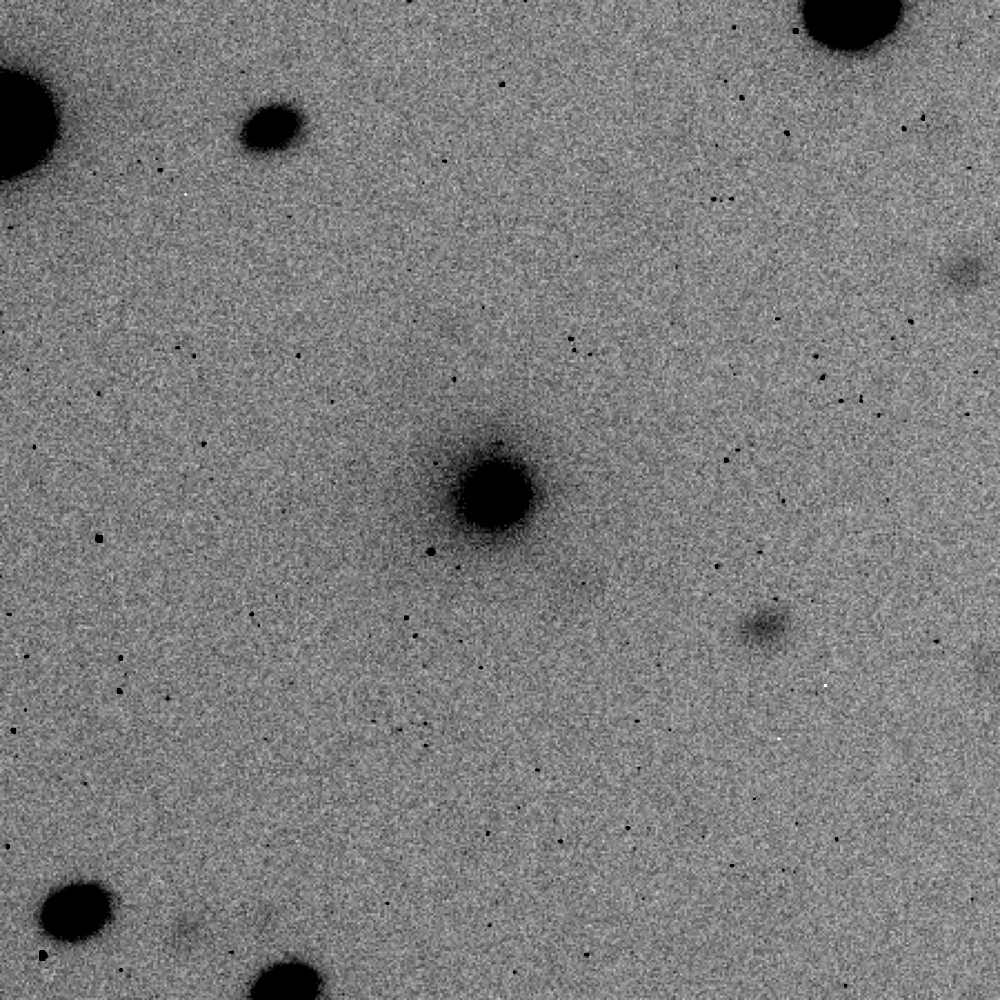}{h}{2024-09-04}{\figsize}{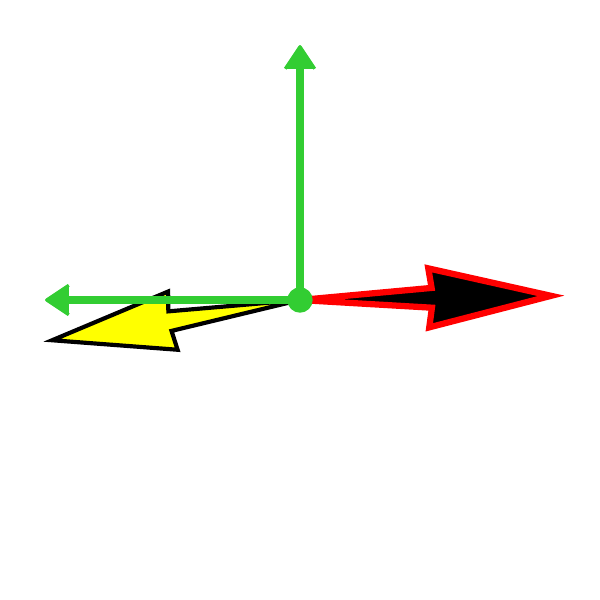}\\
\end{tabular}
\caption{Images of \objname{} (at center) displaying cometary activity 
between the anti-velocity (red arrow) and anti-solar (yellow arrow) directions.
The FOV is 126\arcsec $\times$ 126\arcsec, with North up and East left.
\textbf{(a)} UT 2024 January 12 Lowell Discovery Telescope (LDT) 5$\times$300$s$, \textit{VR}-filter image taken with the Large Monolithic Imager (PI: C. Trujillo, Observers: W. Oldroyd, K. Farrell). 
\textbf{(b)} UT 2024 January 13 Astrophysical Research Consortium telescope imaging camera (ARCTIC) 2$\times$300$s$, \textit{VR}-filter images (Prop. ID 2024B-UW05, PI: C. Chandler, Observers: C. Chandler, W. Oldroyd). 
\textbf{(c)} UT 2024 January 16 Vatican Advanced Technology Telescope 300$s$, V-filter image (PI: C. Trujillo, Observers: C. Chandler, M. Magbanua).
\textbf{(d)} UT 2024 February 15 6$\times$300~s \textit{VR}-band images with ARCTIC (PI C. Chandler, Observers W. Oldroyd, C. Chandler). 
\textbf{(e)} UT 2024 April 11 2$\times$300~s $r$-band ARCTIC images (PI C. Chandler, Observer W. Oldroyd). 
\textbf{(f)} UT 2024 May 9 2$\times$300~s \textit{VR}-band ARCTIC images (PI C. Chandler, Observers W. Oldroyd, C. Chandler). 
\textbf{(g)} UT 2024 August 9 2$\times$300~s \textit{VR}-band ARCTIC images (PI C. Chandler, Observers C. Chandler, M. Frissell, P. Stone). 
\textbf{(h)} UT 2024 September 4 1$\times$300~s \textit{g}-band ARCTIC image (PI C. Chandler, Observers M. Frissell, C. Chandler). 
}
\end{figure*}

Our telescope observations resulted in positive identifications of activity from \objname{}, oriented between the anti-solar and anti-velocity directions, from UT 2024 January 12 to September 4 ($3.035$~au$>r>2.877$~au, $318.1^\circ>\nu<10.5^\circ$; Table \ref{tab:obs}). We conducted telescope observations of \objname{} on the 4.3~m Lowell Discovery Telescope \citep[LDT,][near Happy Jack, Arizona]{2022SPIE12182E..27L}, the 3.5~m Astrophysical Research Consortium telescope \citep[ARC,][Sunspot, New Mexico]{2016SPIE.9908E..5HH}, and the 1.8~m Vatican Advanced Technology Telescope \citep[VATT,][Mount Graham, Arizona]{1997SPIE.2871...74W}. Our first observations as \objname{} approached perihelion, images taken at the LDT on UT 2024 January 12 ($r=3.035$~au and $\nu=318.1^\circ$), showed a field crowded with background stars. However, we detected a linear tail from \objname{}, shown in Figure \ref{fig:obs}(a), that we investigated over the following days with the ARC telescope and the VATT. We captured images with the ARC telescope on UT 2024 January 13 ($r=3.034$~au, $\nu=318.3^\circ$) and February 15 ($r=3.030$~au, $\nu=318.7^\circ$), and these data also indicated a thin tail extending from \objname{}, seen in Figure \ref{fig:obs}(b \& d). The extended tail was not as defined in data from the VATT, though we found faint signs of activity on UT 2024 January 16 ($r=3.029$~au, $\nu=318.9^\circ$; Figure \ref{fig:obs}(c)). Our observations with the ARC telescope on UT 2024 April 11, May 9, August 9, and September 4, indicate more diffuse activity in the form of a coma ($2.877$~au$<r<2.915$~au, $337.3^\circ>\nu<10.5^\circ$; Figure \ref{fig:obs}(e-h)).

For each observation, we noted activity was concentrated between the anti-solar and anti-velocity directions. Given that anti-solar activity is dominated by ionized gas from solar heating and anti-velocity activity is dominated by dust particles, the orientation of activity from 362P is likely a combination of dust and ionized gas. The apparent evolution of tail morphology may be understood by evaluating physical parameters, such as the Sun-Target-Observer (STO) phase angle included in Table \ref{tab:obs}. We find that while the S-T-O angle reaches a minimum (when \objname{} is at opposition and anti-solar activity will appear as a coma) in May 2024, activity appears more diffuse as observations continue past this time.

We identified images of \objname{} in our archival search of DECam \citep{DecamCite}, ZTF \citep{ZTFCITATION}, SkyMapper \citep{Keller2007}, and Las Cumbres Observatory Global Telescope (LCOGT) \citep{2013PASP..125.1031B} data. We confirmed activity near the first epoch in archival DECam, SkyMapper, and LCOGT images, where \objname{} appeared with a coma or diffuse tail (Figure \ref{fig:archival}), all previously unreported activity that extends the known window of activity for that epoch. We also identified activity in numerous archival ZTF images spanning UT 2024 April 3 to UT 2024 May 14 ($2.923$~au$>r>2.888$~au, $335.6^\circ<\nu<344.7^\circ$). We have not specifically assessed the quality of archival data taken outside of perihelion, $83^\circ<\nu<318^\circ$, that didn't result in a detection of activity. This is because the methods, instruments, and circumstances of the archival data collection vary widely among the myriad of data sources. We instead state that the activity we have observed both in archival data and in our new telescope data show activity only near perihelion.

\begin{deluxetable*}{clcrccccc}
\tablenum{1}
\caption{Follow-up telescope observations of activity on \objname{} at the 3.5~m Astrophysical Research Consortium (ARC) telescope, the 4.3~m Lowell Discovery Telescope (LDT), and the 1.8~m Vatican Advanced Technology Telescope (VATT), plus archival observations recovered from the 4~m Blanco Telescope (Blanco), two 2~m Las Cumbres Observatory Global Telescopes (LCOGT), one each from Faulkes North (Hawaii) and Siding Spring (Australia), the 1.35~m SkyMapper telescope at Siding Spring, and the 48'' Samuel Oschin Schmidt telescope (P48) at Palomar. Activity observation sources are abbreviated ``A'' and ''N'' for archival and new telescope observations carried out for this work, respectively. Date, telescope site, true anomaly ($\nu$), heliocentric distance ($r$), filter, exposure number and time, apparent $V$-band magnitude ($m_V$), and Sun Target Observer (S-T-O) Phase Angle for each observed positive detection are included. Quantities for $\nu$, $r$, $m_V$, and S-T-O Phase were retrieved from JPL Horizons on UT 2024 December 3 \citep{giorginiJPLOnLineSolar1996}.}
\label{tab:obs}
\tablehead{
        \colhead{Source} & \colhead{UT Date} & \colhead{Telescope} & \colhead{$\nu$ [deg]} & \colhead{$r$ [au]} & \colhead{Filter(s)} & \colhead{Exposure(s)} & \colhead{$m_V$} & \colhead{S-T-O Phase ($^\circ$)}
    }
\startdata
A & 2015 February 18 & Blanco & 258.9 & 3.865 & \textit{g,r} & 5$\times$86~s + 1$\times$86~s & 18.4 & 4.1\\
A & 2016 March 7 & Blanco & 322.1 & 2.989 & \textit{VR} & 2$\times$250~s & 18.1 & 16.2\\ 
A & 2016 April 9 & Blanco           & 329.2 & 2.942 & \textit{r} & 1$\times$118~s & 17.4 & 7.9\\ 
A & 2016 July 21 & Blanco           & 352.4 & 2.856 & \textit{z} & 1$\times$100~s & 18.3 & 20.6\\ 
A & 2017 July 5  & LCOGT            & 66.3  & 3.281 & \textit{R} & 3$\times$180~s & 18.3 & 12.8\\ 
A & 2017 August 28 & LCOGT          & 75.4  & 3.410 & \textit{R} & 180~s & 18.2 & 6.9\\ 
A & 2017 September 21 & SkyMapper   & 79.2  & 3.469 & \textit{g}, \textit{r} & 1$\times$100~s + 1$\times$100~s & 18.6 & 11.8\\
A & 2017 October 10 & SkyMapper     & 82.2  & 3.516 & \textit{r} & 1$\times$100~s & 18.9 & 14.6\\
A & 2018 September 6 & Blanco  & 123.0  & 4.311 & \textit{g,r} & 1$\times$89~s + 1$\times$45~s & 19.1 & 7.2\\
N & 2024 January 12 & LDT           & 318.1 & 3.035 & \textit{VR} & 5$\times$300~s & 18.9 & 17.6\\
N & 2024 January 13 & ARC           & 318.3 & 3.034 & \textit{VR} & 4$\times$300~s & 18.9 & 17.7\\
N & 2024 January 16 & VATT          & 318.9 & 3.029 & \textit{V}, \textit{R} & 12$\times$120~s + 2$\times$120~s & 18.9 & 18.0\\
N & 2024 February 15 & ARC           & 318.7 & 3.030 & \textit{VR}, \textit{g}, \textit{r}, \textit{i} & 7$\times$300~s & 18.9 & 17.9\\
A & 2024 April 3    & P48           & 335.6 & 2.923 & \textit{g}, \textit{z} & 1$\times$30~s + 1$\times$30~s & 17.7 & 13.0\\ 
N & 2024 April 11   & ARC           & 337.3 & 2.915 & \textit{r} & 2$\times$300~s & 17.5 & 10.8\\
A & 2024 April 30    & P48           & 341.6 & 2.899 & \textit{g}, \textit{z} & 2$\times$30~s + 2$\times$30~s & 17.2 & 5.7\\ 
N & 2024 May 9      & ARC           & 343.6 & 2.892 & \textit{VR} & 2$\times$300~s & 17.1 & 4.9\\
A & 2024 May 14    & P48           & 344.7 & 2.888 & \textit{g}, \textit{z} & 2$\times$30~s + 2$\times$30~s & 17.2 & 5.4\\ 
N & 2024 August 9   & ARC           & 4.6   & 2.868 & \textit{VR} & 2$\times$300~s & 18.4 & 20.6\\
N & 2024 September 4 & ARC           & 10.5   & 2.877 & \textit{g} & 2$\times$300~s & 18.7 & 19.7\\
\enddata
\end{deluxetable*}

\begin{figure*}[ht]
\begin{centering}
\includegraphics[width=0.64\linewidth]{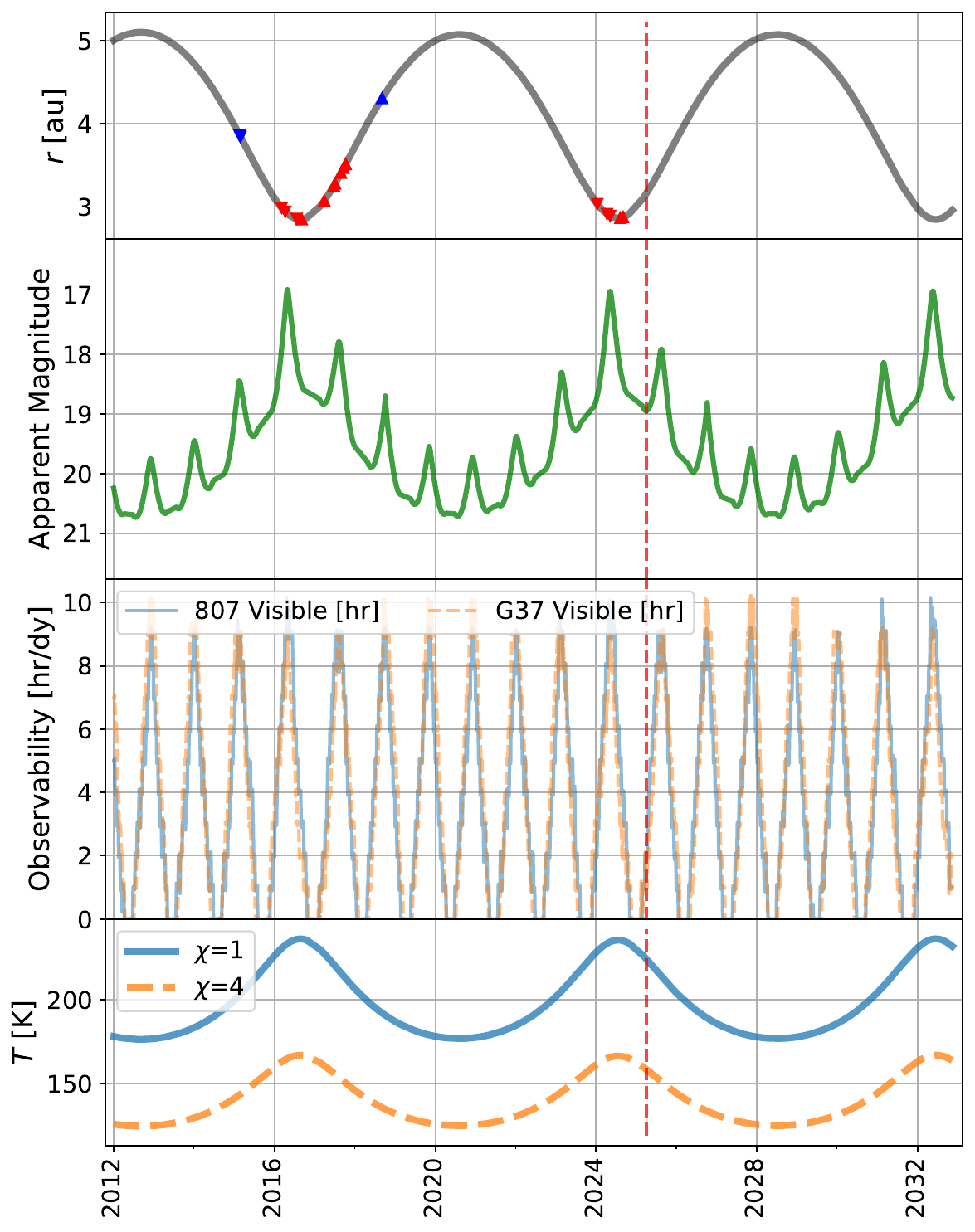}\\
\caption{Observability and Thermodynamical Modeling of \objname{} from 2012 to 2033. Descending, subpanels show heliocentric distance ($r$) in au with markers indicating positive activity detections (red) and negative activity detections (blue), both inbound and outbound (downward and upward pointing triangles, respectively), 
apparent $V$-band magnitude, observability in hours per day, and temperature ($T$) in Kelvin of \objname{} for the thermophysical extremes of $\chi=1$ (flat, Sun-facing slab) and $\chi=4$ (isothermal approximation). 
Activity markers include this work and the observations obtained by \cite{GARCIAMIGANI201812} between 2016 September to 2017 July. 
Our observability metric for Cerro Tololo Inter-American Observatory (CTIO), site code 807 (blue solid line), and the Lowell Discovery Telescope (LDT), site code G37 (orange dashed line), depicts the number of hours \objname{} was observable ($>$15° above the horizon between sunset and sunrise) during a given universal time (UT) observing date. 
The location of \objname{} as of UT 2024 November 23 is noted as a vertical red-dashed line. Heliocentric distance, apparent magnitude, and elevation information for each observatory site were retrieved from JPL on UT 2024 November 23.}
\label{fig:thermo}
\end{centering}
\end{figure*}

\section{Dynamical Analysis}\label{sec:dynamics}
To investigate \objname{}’s dynamical behavior, we ran a suite of dynamical simulations. We created 500 \objname{} clones and used the {\tt IAS15} \citep{2015MNRAS.446.1424R} integrator from the {\tt REBOUND} $N$-body integration package \citep{2012A&A...537A.128R} in {\tt Python} to compute their orbital evolution over $\pm$ 100~kyr. These clones are generated based on the orbital characteristics from JPL Horizons as in \cite{2023ApJ...957L...1O}. We note the Tisserand parameter with respect to Jupiter \citep[$T_J=$ 2.927; retrieved from JPL Horizons on UT 2024 February 28][]{giorginiJPLOnLineSolar1996} falls within the range typical for JFCs \citep[2 $< T_\mathrm{J} <$ 3;][]{1996ASPC..107..173L}.

We investigate the orbital lifetime of \objname{} near the 3:2 interior mean-motion resonance with Jupiter. Our simulations of \objname{} indicate it is currently a JFC in a quasi-Hilda orbit, confirming its current designation \citep{GARCIAMIGANI201812}. Within the past $100$ kyr of our simulations, \objname{} experienced a series of encounters with Jupiter, which perturbed its orbit to that of a JFC. Prior to becoming a JFC, \objname{} was more distant than it is found today, having a 53\% chance of being a Long Period Comet and a 32\% chance of being a Centaur at time of $-100$ kyr. We find only a 15\% chance that \objname{} has remained a JFC over the integrated period of $100$ kyr. Our simulations indicate \objname{} transitioned to its current orbit within the last $20$ kyr, and that over the future $1$ kyr \objname{} experiences orbital chaos, pointing towards its dynamical instability.

\begin{figure*}[ht]
\centering
    \begin{tabular}{cc}
    (a): (153) Hilda Orbit & (b): 362P (quasi-Hilda) Orbit \\
    \includegraphics[width=0.38\linewidth]{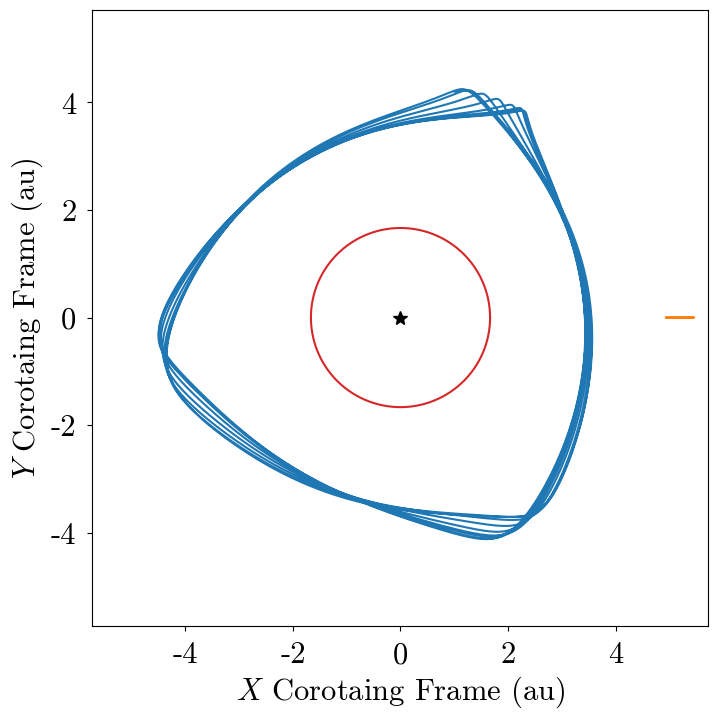} & \includegraphics[width=0.38\linewidth]{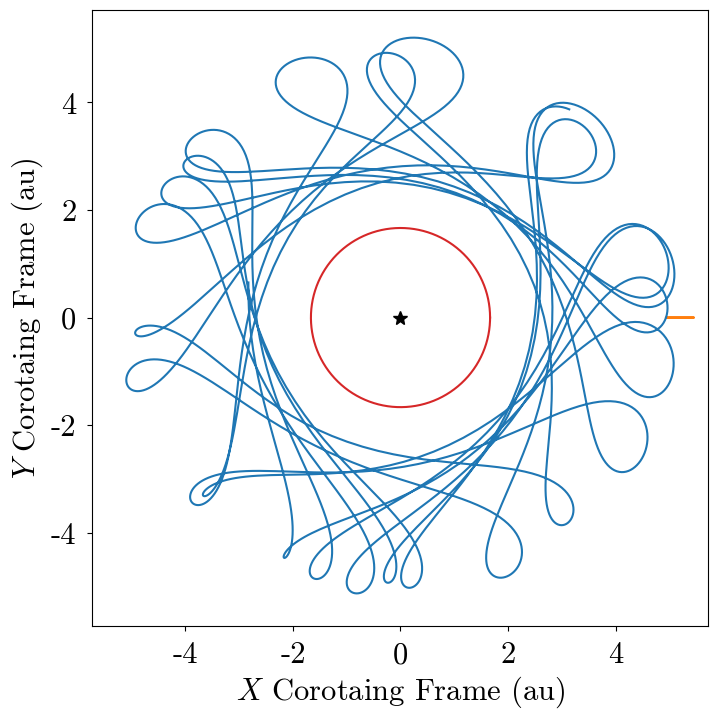} \\
    (c): Orbit Diagram & (d): Semi-Major Axis Evolution \\
    \includegraphics[width=0.38\linewidth]{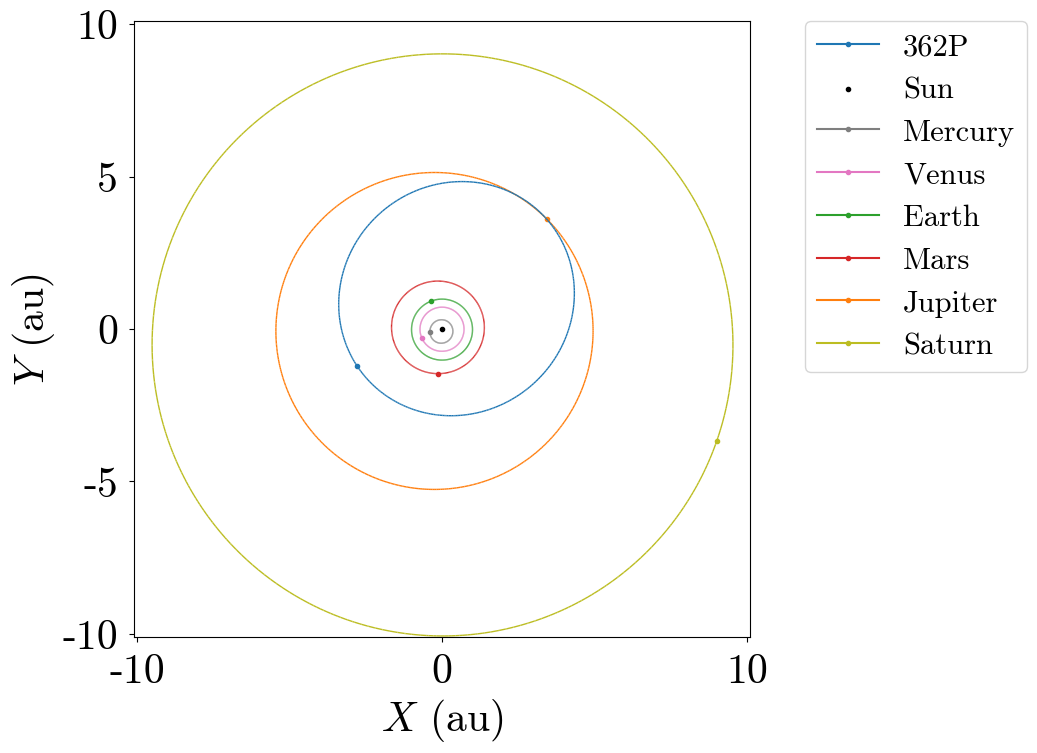} & \includegraphics[width=0.38\linewidth]{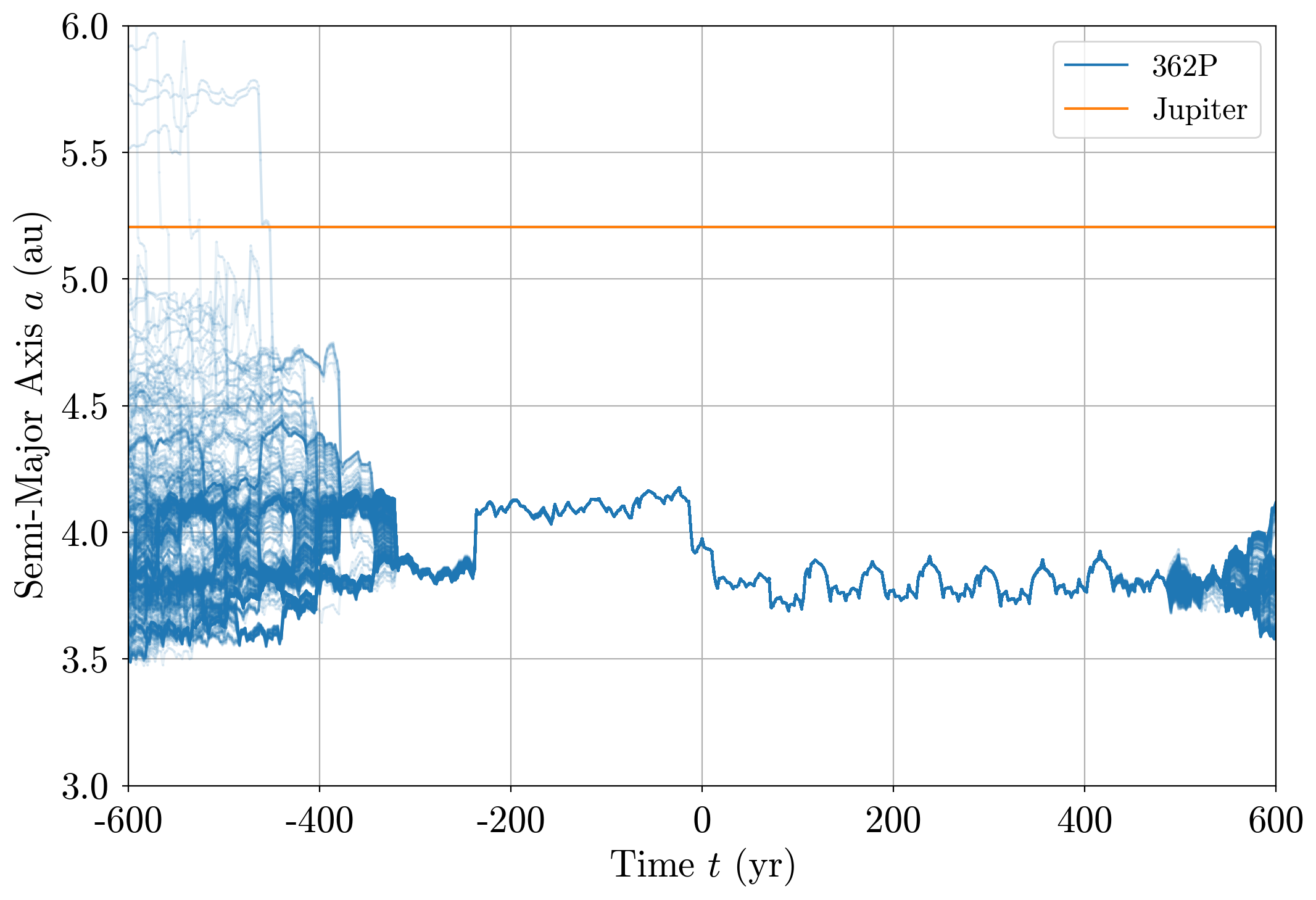} \\
    (e): Heliocentric Distance & (f): Distance to Jupiter \\
    \includegraphics[width=0.38\linewidth]{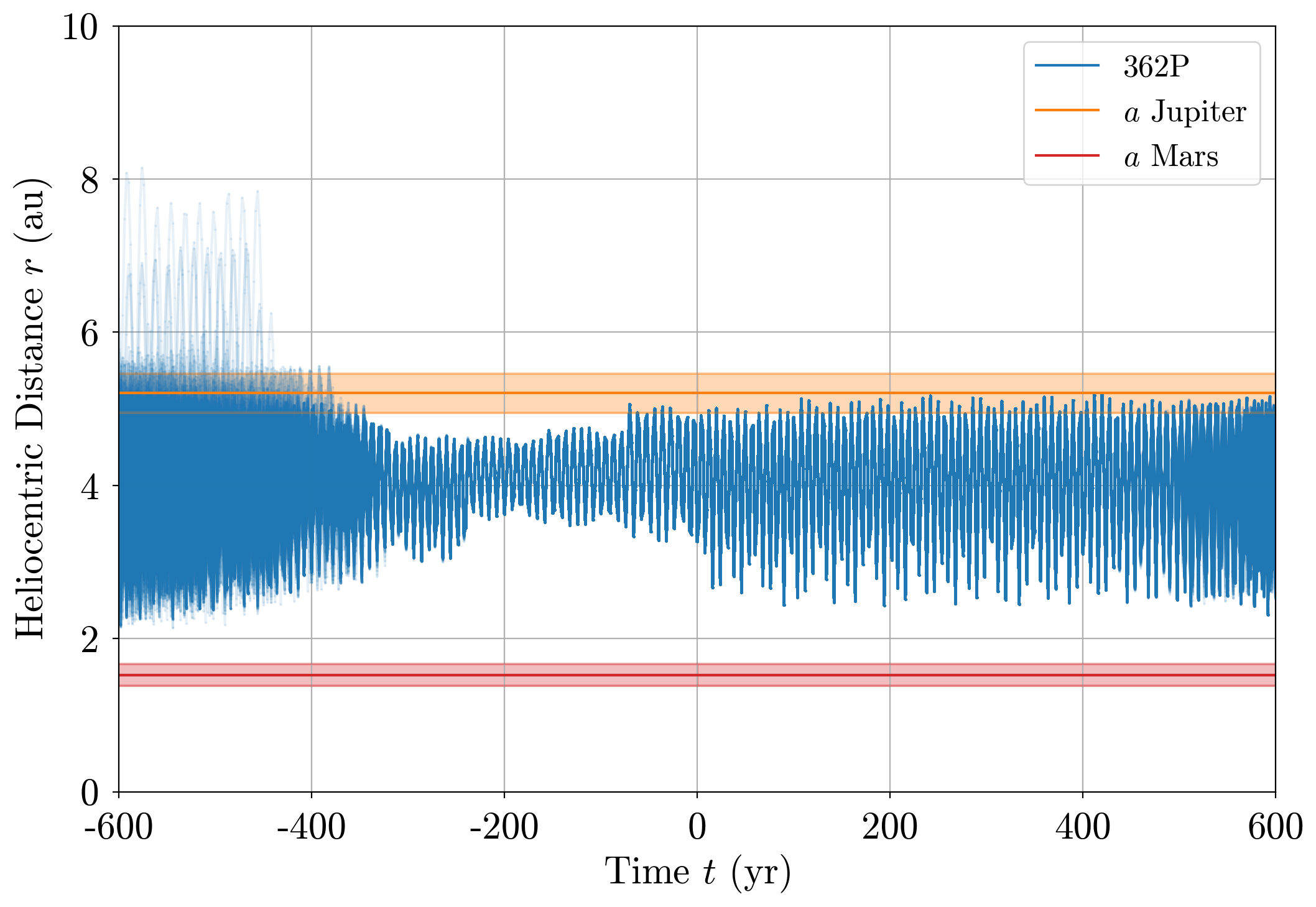} & \includegraphics[width=0.38\linewidth]{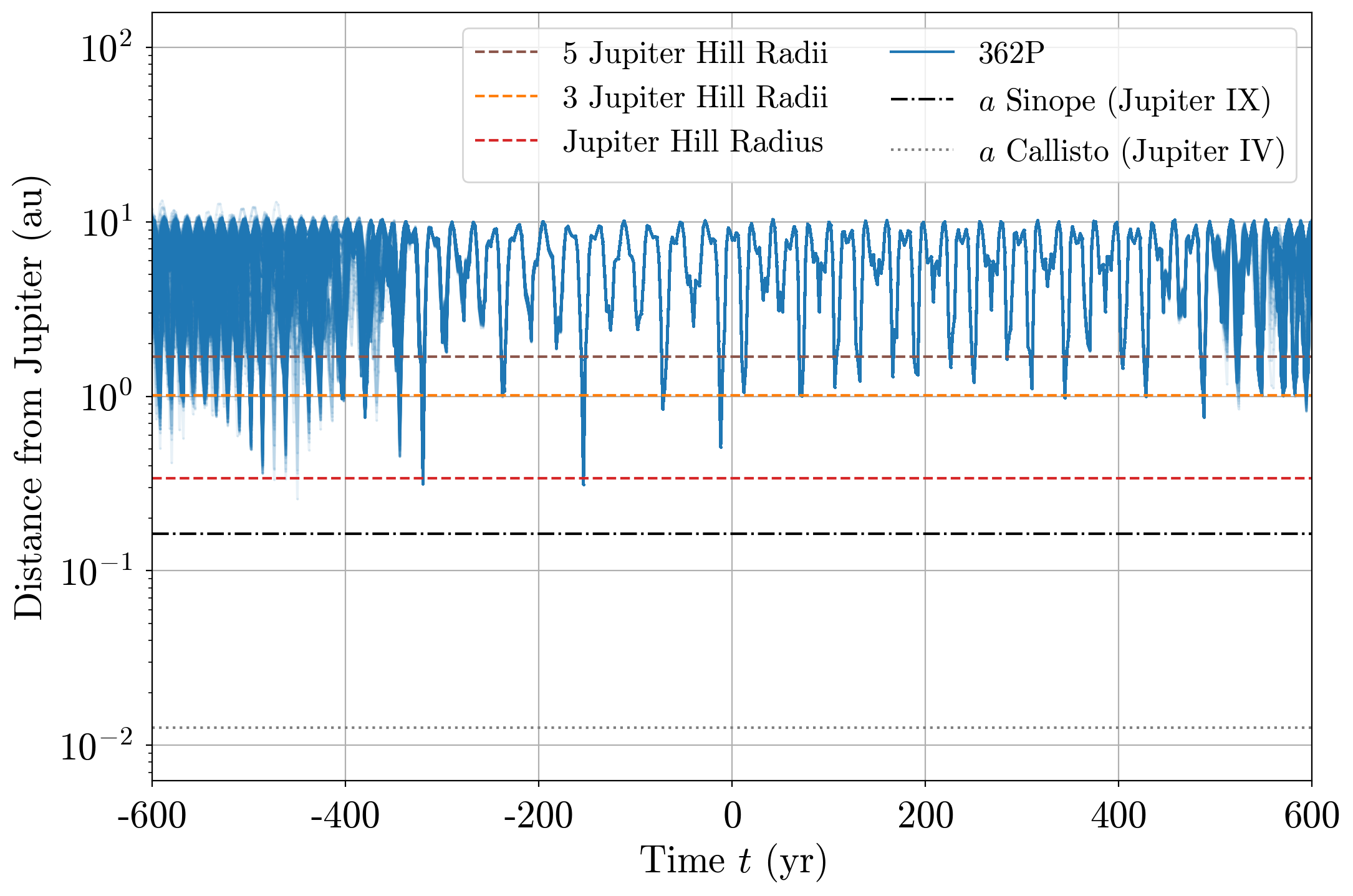} \\
    \end{tabular}
    \caption{(a \& b): Orbits (blue) of (153) Hilda and \objname{}, respectively, in the co-rotating reference frame with Jupiter (orange). The axis "X Corotating frame" indicates the x-axis of the corotating frame, and "Y Corotating frame" indicates the y-axis of the corotating frame. The aphelion distance of Mars (red) is shown for reference. (c): Orbit diagram; note the proximity of the orbits of \objname{} and Jupiter. (d): Semi-major axis evolution of \objname{} orbital clones over $\pm$ 600~yrs; the semi-major axis of Jupiter (orange) is shown for reference. (e): Heliocentric distance evolution of \objname{} clones with orbital distances of Jupiter (orange) and Mars (red) for reference. (f): Logarithmic distance between \objname{} and Jupiter. Horizontal lines representing the depth of close encounters (with fewer Hill radii corresponding to larger gravitational influence), and the orbital distance of two Jovian moons are given for reference.}
\label{fig:dynamics}
\end{figure*}

Figure \ref{fig:dynamics}(a) \& (b) compares the orbit of (153) Hilda, the namesake of the population, to that of \objname{} in a co-rotating frame with Jupiter. Note the triangular pattern followed by (153) Hilda, indicative of the 3:2 interior mean-motion resonance with Jupiter, which protects (153) Hilda (and other objects in this resonance) from close encounters with Jupiter. Since \objname{} is near this resonance, it traces out similar triangular patterns, but the locations of the corners are not well-confined. This allows \objname{} to experience frequent (9 within 5 Jupiter Hill radii in the timespan $\pm200$~yr) close encounters with Jupiter that destabilize its orbit and indicate the comet is a quasi-Hilda.

The orientation indicates \objname{} may interact with Jupiter, given the proximity of their orbits (\ref{fig:dynamics}(c)). We note that \objname{} has an orbital inclination of $15.556^\circ$, and JPL Horizons \citep{giorginiJPLOnLineSolar1996} reports the minimum orbital intersection distance (MOID) with Jupiter is $0.354$~au. Encounters with Jupiter will perturb the orbit of \objname{} and could result in changes to its orbital parameters and eventually its orbital class. Figure \ref{fig:dynamics}(d) shows changes in semi-major axis that the \objname{} orbital clones experience. While the semi-major axis experiences chaos near $\pm1$~kyr, within $\pm600$~yr we see regular, short-term changes to this parameter. Orbital changes on both of these timescales are caused by close encounters between \objname{} and Jupiter (Figure \ref{fig:dynamics}(f)). While these simulations do not ensure the future dynamics of \objname{}, they explore the statistical likelihood of its dynamical migration outside of its current quasi-Hilda orbit. 

The orbital evolution of \objname{}, given close encounters with Jupiter, may provide an explanation for negative activity around the discovery of the object in 2008 ($r=3.246$~au, $\nu=332.3^\circ$). Our simulations support that \objname{} did have a recent close encounter, within 3 Jupiter Hill Radii, with that altered both the semi-major axis (Figure \ref{fig:dynamics}(d)) and heliocentric distance (Figure \ref{fig:dynamics}(e)). Following the encounter, the shorter perihelion distance of \objname{} allowed sublimation to occur near perihelion in 2016/2017 and 2024. This finding is also supported by JPL Horizons \citep{giorginiJPLOnLineSolar1996}, where the most recent close encounter of \objname{} with Jupiter occurred during September 2011, after the object was first discovered and not identified to be active.

Given our findings that interactions with Jupiter are frequent ($\sim1$ per $40$~yr), \objname{} will experience orbital chaos beyond 1~kyr. From forward integration of \objname{} clones, we find it will remain both a quasi-Hilda object and a JFC into the chaotic period. Notably, chaos dominates before $\sim-300$~yr and after $\sim500$~yr in Figure \ref{fig:dynamics}(d)-(f). The semi-major axis of \objname{} presents an interesting investigation, given the recent shift to a lower semi-major axis (Figure \ref{fig:dynamics}(d)). Similar drops in semi-major axis may have a positive relationship to activity \citep{lilly2023semimajor}.

\section{Activity Assessment} \label{sec:activityassessment}
We investigate the observability and thermodynamics of \objname{} with the aid of Figure \ref{fig:thermo}. The figure contains four panels that share a temporal horizontal axis. This figure helps identify patterns in activity identification that may correlate to different factors, specifically heliocentric distance, apparent $V$-band magnitude, how ``observable'' an object was each night by an observatory in the Northern or Southern hemisphere, and the surface temperature. The temperature is derived through simple thermodynamical modeling, described in detail in \cite{chandlerCometaryActivityDiscovered2020b} and \cite{HSIEH2015289}. In short, we compute the surface temperature for a gray, airless body at a given heliocentric distance from the Sun, informed by the body's rotation rate, bound by the extremes of $\chi=1$ (a flat, Sun-facing slab) and $\chi=4$ (a rapidly rotating, isothermal body).

We find as \objname{} approaches perihelion, it sees a local minimum in heliocentric distance ($r$), and local maxima in apparent magnitude ($m$), observability, and temperature ($T$). Coincident peaks in apparent magnitude and observability indicate opposition events, and coincident troughs mark conjunctions. This increase in temperature is concurrent with the approach to perihelion, indicating increased solar heating may be responsible for the observed activity. Moreover, our thermal modeling shows that \objname{} currently experiences $\pm25$~K temperature variations which bring or keep it well into the regime where water sublimation would occur.

We cataloged the known images of \objname{} activity, from literature and this work, and marked these inbound and outbound data points on the distance plot (Figure \ref{fig:thermo}). An important element of our archival search is recognizing that exposure time, filter, and aperture size vary across facilities. We note that, while our archival search did not yield positive activity detections from \objname{} between $83^\circ<\nu<318^\circ$, deeper images may reveal activity during this period that was not visible in archival images. In other words, while there is not evidence of activity during these times, we cannot rule out the possibility that activity went undetected.

In the first activity epoch, for which most activity observations occurred post-perihelion, activity appeared as a diffuse coma (Figure \ref{fig:archival}) near perihelion but evolved into a linear tail during the outgoing arc \citep{GARCIAMIGANI201812}. In our discovery of the second activity epoch, as \objname{} approached perihelion, activity first appeared as a linear tail extending from the comet's nucleus (Figure \ref{fig:obs}(a), (b), \& (c)) with a position angle between the anti-solar and anti-velocity directions. Later, the tail became both brighter and more diffuse (Figure \ref{fig:obs}(e), (f)). The most recent (and post-perihelion) images show little to no distinct tail but a more dispersed dust cloud condensed around the nucleus.

Both activity epochs (2016 -- 2017 and 2024) occurring near perihelion establishes a pattern of recurrent activity near perihelion. Finding activity localized near perihelion suggests volatile sublimation driven by increased solar heating. Additionally, the circumstances of repeated activity near perihelion indicate that continued follow-up telescope observations of \objname{} would be useful to further constrain the activity pattern. Given the changes to the orbit of \objname{} (Figure \ref{fig:dynamics}(d \& e)), we expect the established activity pattern to continue. Further telescope observations can measure how the activity of \objname{} evolves further into the 2024 July perihelion passage for a comparison of orbital space comparable to the 2016--2017 apparition, as well as further investigate the connection between dynamics and its potential for activity.

\section{Summary}\label{sec:sum}
\objname{} is a Jupiter Family Comet (JFC) in a quasi-Hilda orbit with a single previously known activity epoch between 2016 September and 2017 June \citep[][]{GARCIAMIGANI201812}. Our discovery of a second activity epoch for \objname{}, between UT 2024 January 12 to September 4 ($3.035$~au$>r>2.877$~au, $318.1^\circ<\nu>10.5^\circ$; Figure \ref{fig:obs}, indicates repeated sublimation-driven activity near perihelion (Figure \ref{fig:thermo}) and, therefore, supports the presence of volatile materials (Section \ref{sec:activityassessment}).

Our observations (Table \ref{tab:obs}) from the Astrophysical Research Consortium (ARC) telescope, Lowell Discovery Telescope (LDT), and Vatican Advanced Technology Telescope (VATT) contributed to our successful discovery of repeated activity. These observations were taken as \objname{} approached perihelion and extended into the post perihelion arc ($3.035$~au$>r>2.877$~au, $318.1^\circ<\nu>10.5^\circ$ Table \ref{tab:obs}). During the observations reported in \cite{GARCIAMIGANI201812} and the first activity detection ($2.850$~au$<r<3.323$~au, $0.0^\circ<\nu<69.9^\circ$), activity evolved from a coma to a thin linear tail. In our observations, we identify an extended tail from \objname{} over the incoming perihelion arc ($318.1^\circ<\nu< 318.9^\circ$) that subsequently evolved into a diffuse, comae-like cloud post-perihelion ($4.6^\circ<\nu<10.5^\circ$). Our archival search confirmed activity during the first activity epoch with DECam images from UT 2016 March 7, April 9, and July 21, LCOGT archival images taken on UT 2017 July 5 and UT 2017 August 28 at Faulkes North  (Hawaii) and Siding Spring (Australia), respectively, and SkyMapper images from UT 2017 September 21 and October 10. Our archival discoveries extend the first activity apparition timeline to between UT 2016 March 7 and UT 2017 October 10. Following our discovery of the new activity apparition, we identified numerous images of the current activity apparition in ZTF data spanning at least UT 2024 April 3 to 2024 May 14.

We confirm that \objname{} is a QHC, and that the object will experience dynamical chaos beyond $1$~kyr, limiting our understanding of the dynamical future for \objname{}. Interactions with Jupiter will influence its dynamical class as well as semi-major axis distance, potentially supporting observations for a third epoch of activity near the subsequent projected perihelion passage on UT 2031 June 14. Our investigation into the orbital past of \objname{} indicates a high potential that it has migrated from a Long Period Comet orbit ($\sim$53\%) or a Centaur orbit ($\sim$32\%), though it could have remained a JFC ($\sim$15\%) during the past $100$~kyr. In summary, the most likely case is that \objname{} has migrated to a closer orbit to the Sun in the last $\sim300$~yrs, and in so doing, this increase in solar heating triggered the recent perihelion activity in two epochs. We conclude this activity is volatile-driven. We predict that \objname{} will remain repeatedly active in future perihelion passages and that it will experience another chaotic period in about $1$~kyr.

\objname{} passed perihelion on UT 2024 July 20, thus additional observations acquired hereafter will parallel the same orbital region as was extensively observed during the 2016--2017 apparition. These timely observations are essential for understanding the story of \objname{}, especially to further characterize the sublimation of volatile ices, describe the activity morphology evolution, bolster the diagnosis of sublimation as the dominant activity mechanism, and constrain the true anomaly range for which \objname{} is found to be active.  Based on previous cometary activity from \objname{} continuing well after perihelion, we recommend observations from the community through at least 2024 December. We also recommend that observations continue to the comet's following perihelion passage on UT 2031 June 14, with such observations pinpointing the true anomaly activity range and supporting the diagnoses of thermochemistry and volatile species on \objname{}.



\section*{Acknowledgements}

A special thank you to the Reviewer, whose comments greatly improved the quality of this manuscript.

This work was funded in part by NASA grant 80NSSC21K0114 (W.J.O., C.A.T., and K.A.F.). This work was made possible in part through the State of Arizona Technology \& Research Initiative Program. K.A.F. and C.A.T.\ acknowledge support from the Northern Arizona University College of Environment, Forestry and Natural Sciences. C.O.C. and C.A.T.\ acknowledge support from the NASA Solar System Observations program (grant 80NSSC19K0869). This material is based upon work supported by the NSF Graduate Research Fellowship Program under grant No.\ 2018258765 and grant No.\ 2020303693. Any opinions, findings, and conclusions or recommendations expressed in this material are those of the author(s) and do not necessarily reflect the views of the NSF.

We thank Arthur and Jeanie Chandler for their ongoing support. 
This research received support through Schmidt Sciences. 
Chandler and Frissell acknowledge support from the DiRAC Institute in the Department of Astronomy at the University of Washington. The DiRAC Institute is supported through generous gifts from the Charles and Lisa Simonyi Fund for Arts and Sciences, and the Washington Research Foundation. We thank Jessica Birky (UW), Tobin Wainer, Eric Bellm, and David Wang (UW) for contributing telescope time at APO.

Computational analyses were run on Northern Arizona University’s Monsoon computing cluster, funded by Arizona’s Technology and Research Initiative Fund. We thank Chris Coffey (NAU) and the NAU High Performance Computing Support team, who have made this work possible. This research has made use of NASA’s Astrophysics Data System. This research has made use of data and/or services provided by the International Astronomical Union's Minor Planet Center. This research has made use of data and services provided by JPL Horizons \citep{giorginiJPLOnLineSolar1996}. This work made use of AstOrb, the Lowell Observatory Asteroid Orbit Database \textit{astorbDB} \citep{1994IAUS..160..477B,moskovitzAstorbDatabaseLowell2021}. World Coordinate System corrections were facilitated by \textit{Astrometry.net} \citep{langAstrometryNetBlind2010}. This research has made use of The Institut de M\'ecanique C\'eleste et de Calcul des \'Eph\'em\'erides SkyBoT Virtual Observatory tool \citep{berthierSkyBoTNewVO2006} and 
the Canadian Astronomical Data Center (CADC) Solar System Object Information Search \citep{gwyn2012ssos}. 
World Coordinate System (WCS) corrections facilitated by the \textit{Astrometry.net} software suite \citep{langAstrometryNetBlind2010}.

Based on observations obtained with the Apache Point Observatory 3.5-meter telescope, which is owned and operated by the Astrophysical Research Consortium. 
Observations made use of Astrophysical Research Consortium Telescope Imaging Camera (ARCTIC) imager \citep{2016SPIE.9908E..5HH}. ARCTIC data reduction made use of the \texttt{acronym} software package \citep{weisenburger2017acronym}. 

The VATT referenced herein refers to the Vatican Observatory’s Alice P. Lennon Telescope and Thomas J. Bannan Astrophysics Facility. We are grateful to the Vatican Observatory for the generous time allocations (Proposal ID S165, PIs Chandler and Oldroyd). A special thanks to Vatican Observatory Director Br. Guy Consolmagno, S.J. Vice Director for Tucson Vatican Observatory Research Group Rev.~Pavel Gabor, S.J., Telescope Scientist Rev. Richard P. Boyle, S.J., Gary Gray, Chris Johnson, Michael Franz, and Summer Franks. 
These results made use of the Lowell Discovery Telescope (LDT) at Lowell Observatory.  Lowell is a private, non-profit institution dedicated to astrophysical research and public appreciation of astronomy and operates the LDT in partnership with Boston University, the University of Maryland, the University of Toledo, Northern Arizona University and Yale University. The Large Monolithic Imager was built by Lowell Observatory using funds provided by the National Science Foundation (AST-1005313). This work makes use of observations from the Las Cumbres Observatory global telescope networks Faulkes North Telescope (Hawaii) and Siding Spring (Australia).
SkyMapper \citep{Keller2007} is owned and operated by The Australian National University's Research School of Astronomy and Astrophysics. The SkyMapper Southern Survey has been funded in part through ARC LIEF grant LE130100104 from the Australian Research Council, awarded to the University of Sydney, the Australian National University, Swinburne University of Technology, the University of Queensland, the University of Western Australia, the University of Melbourne, Curtin University of Technology, Monash University and the Australian Astronomical Observatory. The SkyMapper Southern Survey dataset has been produced with the support of the National Computational Infrastructure (NCI) in Canberra, Australia.

This project used data obtained with the Dark Energy Camera (DECam), which was constructed by the Dark Energy Survey (DES) collaboration. Funding for the DES Projects has been provided by the U.S. Department of Energy, the U.S. National Science Foundation, the Ministry of Science and Education of Spain, the Science and Technology Facilities Council of the United Kingdom, the Higher Education Funding Council for England, the National Center for Supercomputing Applications at the University of Illinois at Urbana-Champaign, the Kavli Institute of Cosmological Physics at the University of Chicago, Center for Cosmology and Astro-Particle Physics at the Ohio State University, the Mitchell Institute for Fundamental Physics and Astronomy at Texas A\&M University, Financiadora de Estudos e Projetos, Funda\c{c}\~{a}o Carlos Chagas Filho de Amparo, Financiadora de Estudos e Projetos, Funda\c{c}\~ao Carlos Chagas Filho de Amparo \`{a} Pesquisa do Estado do Rio de Janeiro, Conselho Nacional de Desenvolvimento Cient\'{i}fico e Tecnol\'{o}gico and the Minist\'{e}rio da Ci\^{e}ncia, Tecnologia e Inova\c{c}\~{a}o, the Deutsche Forschungsgemeinschaft and the Collaborating Institutions in the Dark Energy Survey. The Collaborating Institutions are Argonne National Laboratory, the University of California at Santa Cruz, the University of Cambridge, Centro de Investigaciones En\'{e}rgeticas, Medioambientales y Tecnol\'{o}gicas–Madrid, the University of Chicago, University College London, the DES-Brazil Consortium, the University of Edinburgh, the Eidgen\"ossische Technische Hochschule (ETH) Z\"urich, Fermi National Accelerator Laboratory, the University of Illinois at Urbana-Champaign, the Institut de Ci\`{e}ncies de l'Espai (IEEC/CSIC), the Institut de Física d'Altes Energies, Lawrence Berkeley National Laboratory, the Ludwig-Maximilians Universit\"{a}t M\"{u}nchen and the associated Excellence Cluster Universe, the University of Michigan, the National Optical Astronomy Observatory, the University of Nottingham, the Ohio State University, the University of Pennsylvania, the University of Portsmouth, SLAC National Accelerator Laboratory, Stanford University, the University of Sussex, and Texas A\&M University.

Based on observations at Cerro Tololo Inter-American Observatory, National Optical Astronomy Observatory (NOAO Prop. ID 2016A-0189, PI: Rest; NOAO Prop. ID 2014B-0404, PI: Schlegel; NOAO Prop. ID 2016A-0190, PI: Dey), which is operated by the Association of Universities for Research in Astronomy (AURA) under a cooperative agreement with the National Science Foundation. This research has made use of the NASA/IPAC Infrared Science Archive, which is funded by the National Aeronautics and Space Administration and operated by the California Institute of Technology.

Based on observations obtained with the Samuel Oschin 48-inch Telescope at the Palomar Observatory as part of the Zwicky Transient Facility project. ZTF is supported by the National Science Foundation under Grant No. AST-1440341 and a collaboration including Caltech, IPAC, the Weizmann Institute for Science, the Oskar Klein Center at Stockholm University, the University of Maryland, the University of Washington, Deutsches Elektronen-Synchrotron and Humboldt University, Los Alamos National Laboratories, the TANGO Consortium of Taiwan, the University of Wisconsin at Milwaukee, and Lawrence Berkeley National Laboratories. Operations are conducted by COO, IPAC, and UW.

Based on observations obtained with the Samuel Oschin Telescope 48-inch and the 60-inch Telescope at the Palomar Observatory as part of the Zwicky Transient Facility project. ZTF is supported by the National Science Foundation under Grant No. AST-2034437 and a collaboration including Caltech, IPAC, the Weizmann Institute for Science, the Oskar Klein Center at Stockholm University, the University of Maryland, Deutsches Elektronen-Synchrotron and Humboldt University, the TANGO Consortium of Taiwan, the University of Wisconsin at Milwaukee, Trinity College Dublin, Lawrence Livermore National Laboratories, and IN2P3, France. Operations are conducted by COO, IPAC, and UW.

\vspace{8pt}
\facilities{
ARC: 3.5~m (ARCTIC), 
CTIO:4m (DECam), 
IRSA (IPAC; \citealt{2006ASPC..351..367B})\footnote{\url{https://www.ipac.caltech.edu/doi/irsa/10.26131/IRSA539}}, 
LDT: 4.3~m (LMI),
MGIO: 1.8~m (VATT), 
PO:1.2 m (ZTF; \citealt{2019PASP..131a8002B}), 
SkyMapper: 1.3~m.
}

\software{
        {\tt acronym} \citep{2017JOSS....2..102L},
        {\tt astrometry.net} \citep{langAstrometryNetBlind2010}, 
        {\tt astropy} \citep{robitailleAstropyCommunityPython2013}, 
        {\tt Matplotlib} \citep{hunterMatplotlib2DGraphics2007},
        {\tt NumPy} \citep{harrisArrayProgrammingNumPy2020},
        {\tt pandas} \citep{rebackPandasdevPandasPandas2022},
        {\tt Photutils} \citep{larry_bradley_2023_7946442},
        {\tt REBOUND} \citep{2012A&A...537A.128R,2015MNRAS.446.1424R},
        {\tt SAOImageDS9} \citep{joyeNewFeaturesSAOImage2006},
        {\tt SciPy} \citep{virtanenSciPyFundamentalAlgorithms2020},
        \texttt{tqdm} \citep{casper_da_costa_luis_2020_4293724},
        \texttt{VizieR} \citep{ochsenbeinVizieRDatabaseAstronomical2000}.
          }
\appendix

\section{Archival Activity Observations}
\newcommand{\thisfigsize}{0.3}
\begin{figure*}
    \centering
    \begin{tabular}{ccc}
        \labelpicA{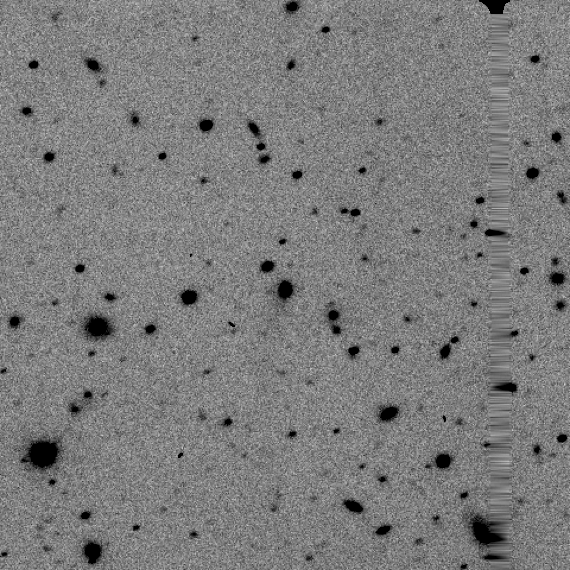}{a}{2016-03-07}{\thisfigsize}{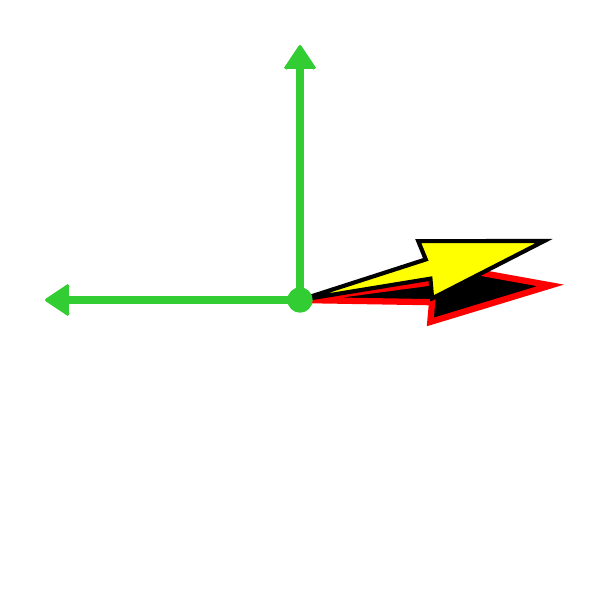} & 
        \labelpicA{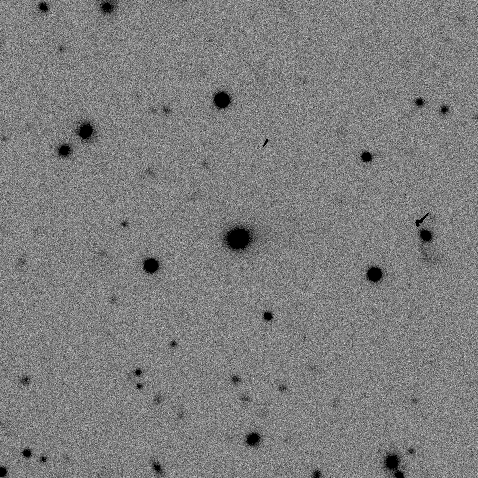}{b}{2016-04-09}{\thisfigsize}{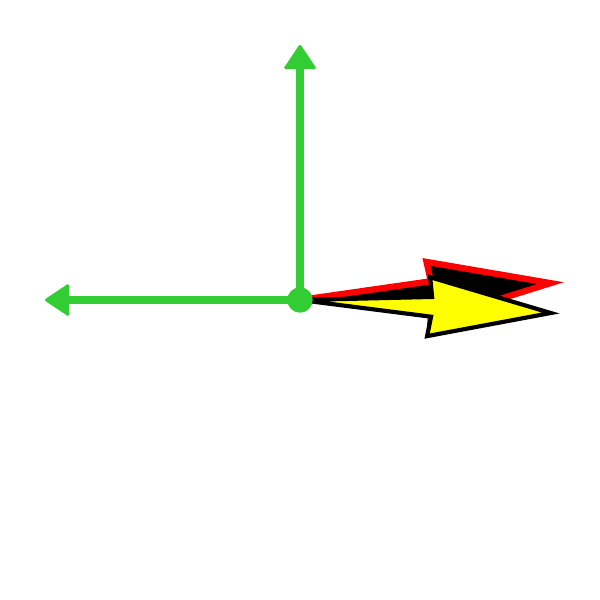} & 
        \labelpicA{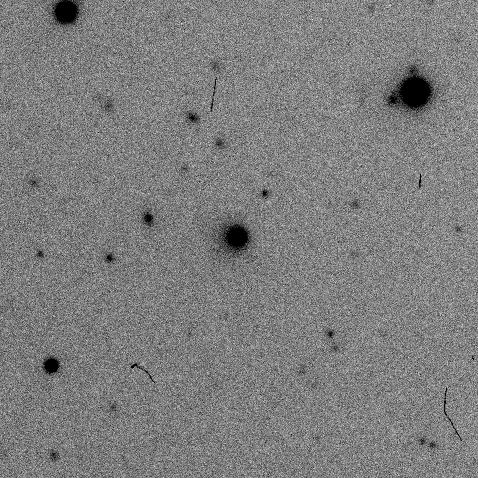}{c}{2016-07-21}{\thisfigsize}{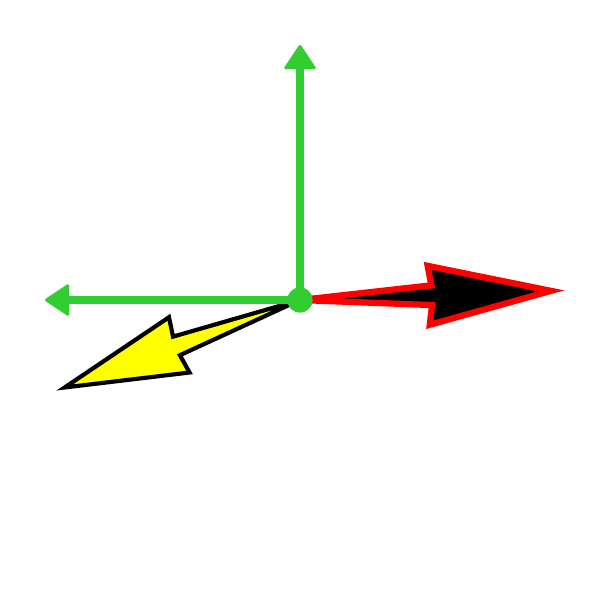} 
        \\ 
        \labelpicA{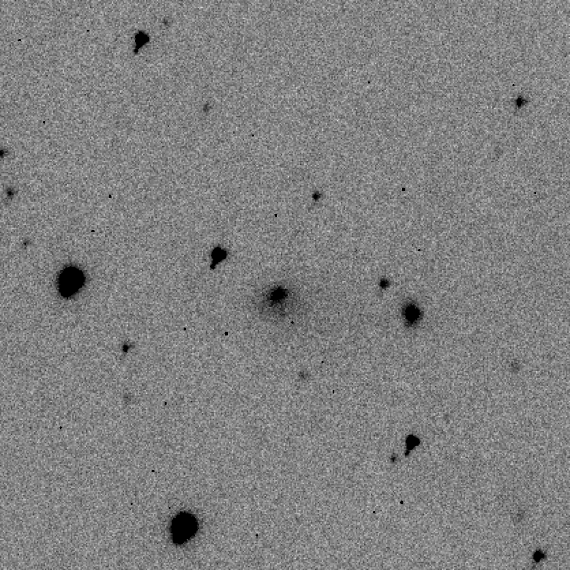}{d}{2017-07-05}{\thisfigsize}{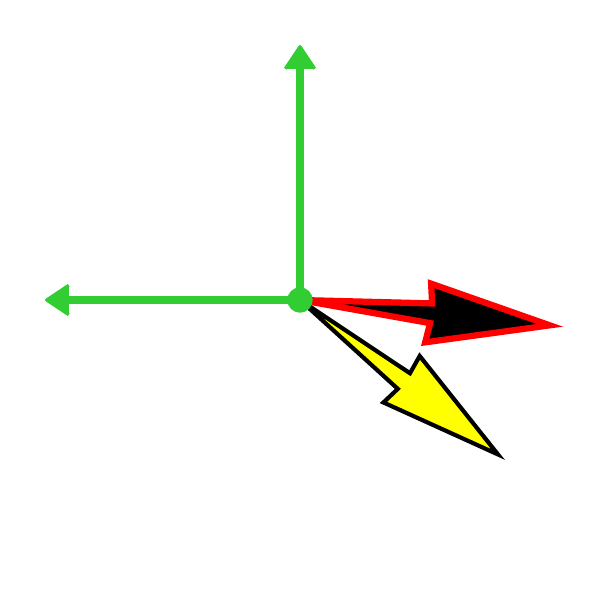}& 
        \labelpicA{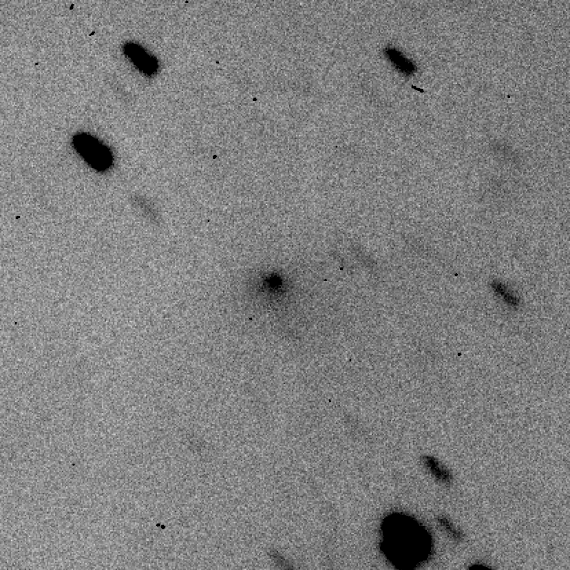}{e}{2017-08-28}
        {\thisfigsize}{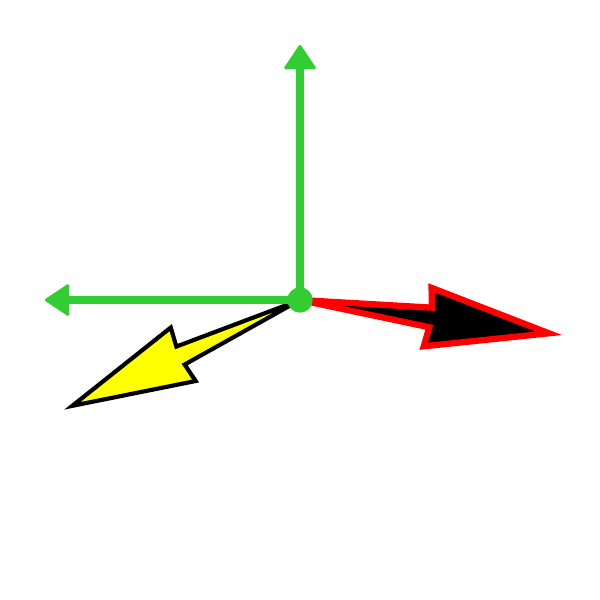}&
        \labelpicA{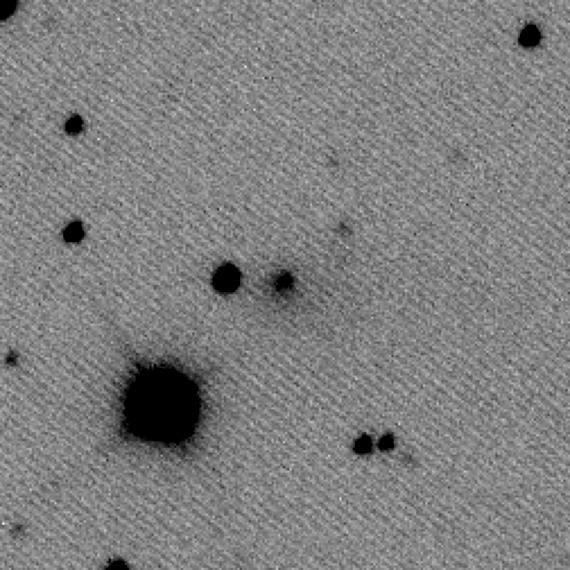}{f}{2017-09-21}{\thisfigsize}{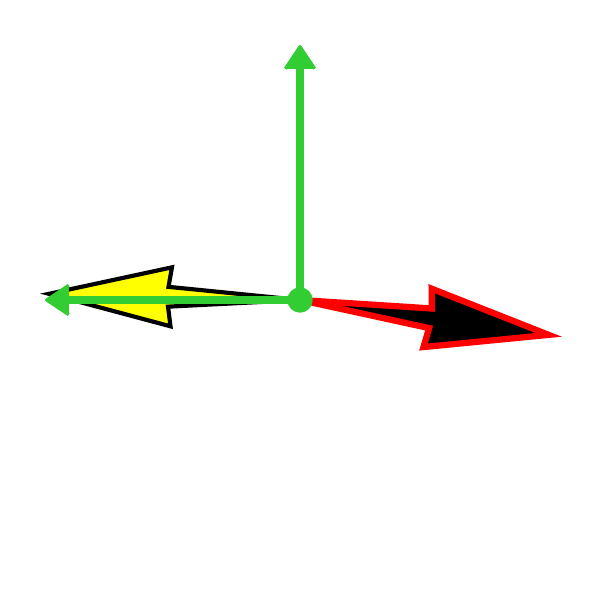}
        \\
        \labelpicA{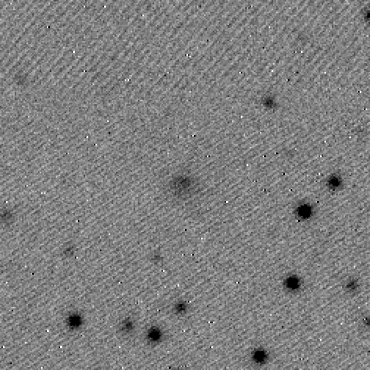}{g}{2017-10-10}{\thisfigsize}{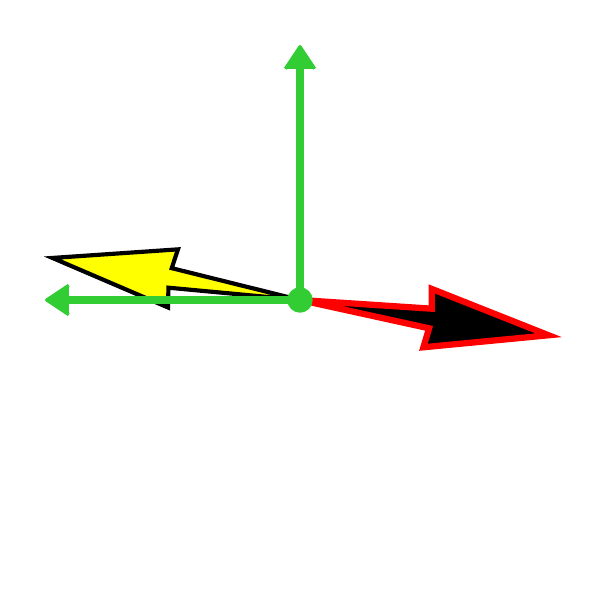}&
        \labelpicA{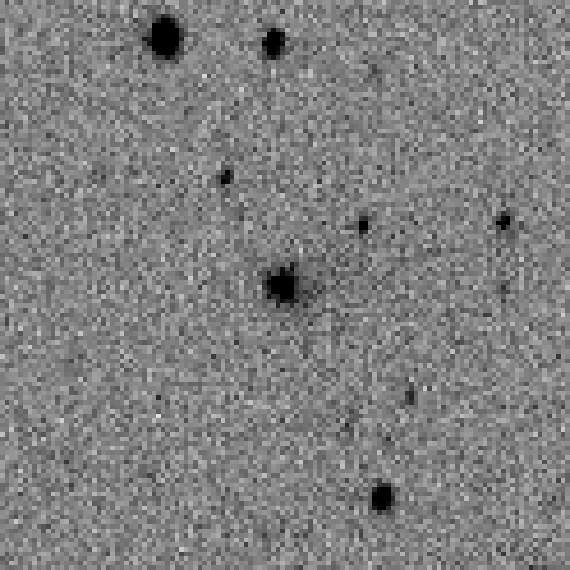}{h}{2024-04-03}{\thisfigsize}{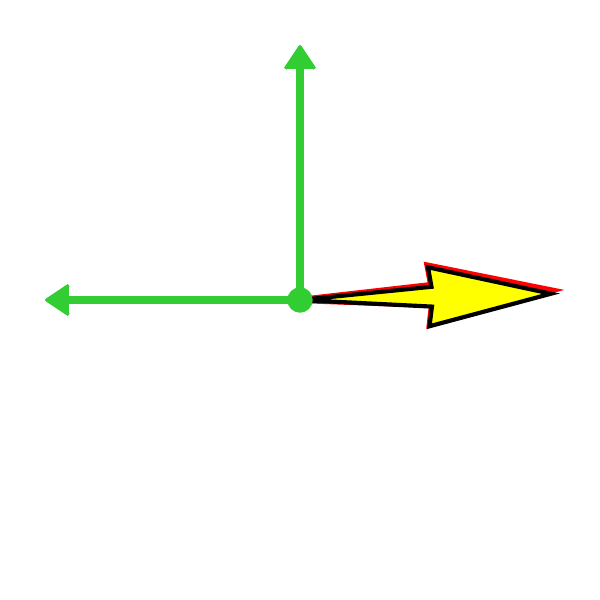} & 
        \labelpicA{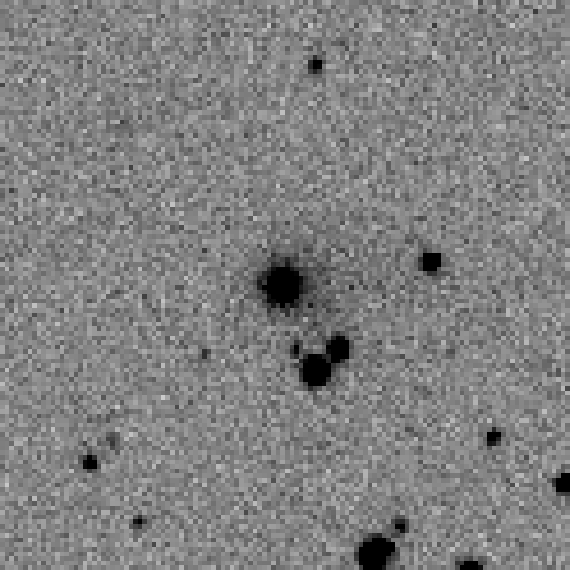}{i}{2024-04-30}{\thisfigsize}{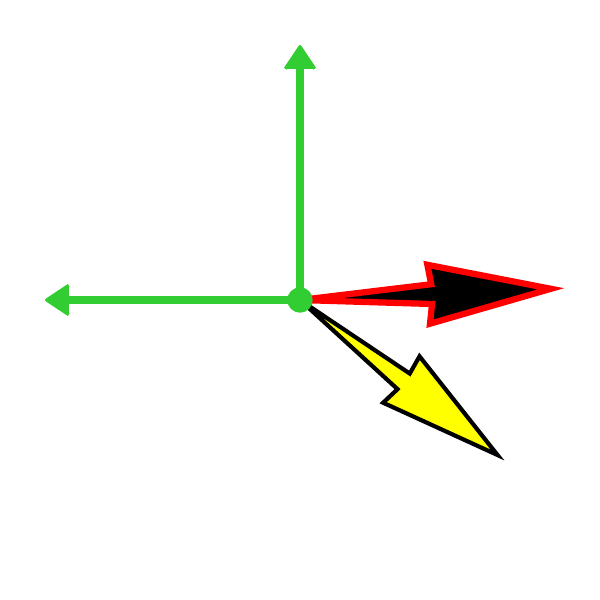}
        \\
        \labelpicA{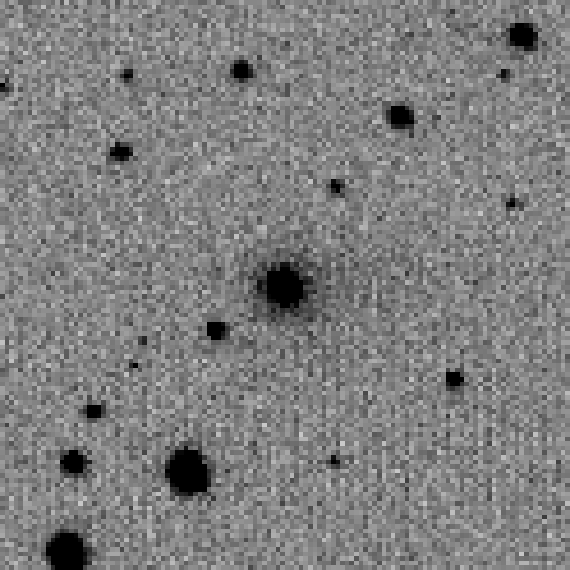}{j}{2024-05-14}{\thisfigsize}{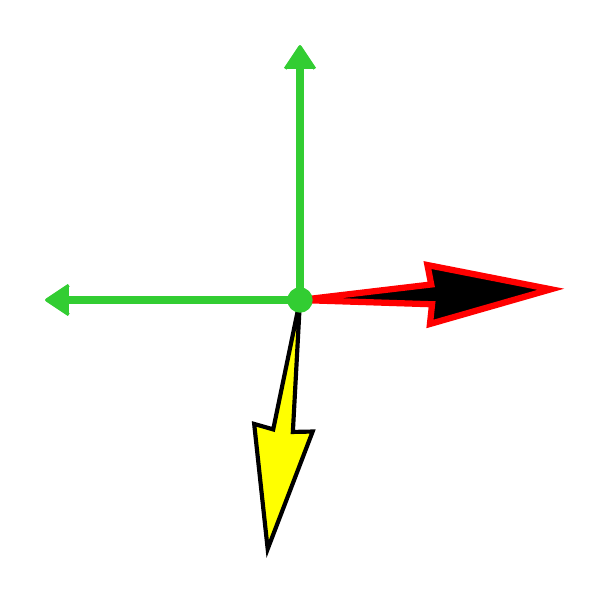}
        \\
    \end{tabular}
    \caption{Caption continues on next page.}
    \label{fig:archival}
\end{figure*}

\setcounter{figure}{3}
\begin{figure}
    \caption{For all images, \objname{} is at the center of these 126\arcsec$\times$126\arcsec fields, with North up and East left. Anti-solar (yellow arrow) and anti-motion (red-outlined black arrow) on-sky directions are indicated. 
    \textbf{(a)} UT 2016 March 7 co-added 2$\times$150~s \textit{VR}-band DECam images (Prop. ID 2016A-0189, PI Rest, observers Armin Rest, DJJ). 
    \textbf{(b)} UT 2016 April 9 1$\times$118~s $r$-band DECam exposure (Prop. ID 2014B-0404, PI Schlegel, observers Arjun Dey, David James). 
    \textbf{(c)} UT 2016 July 21 1$\times$211~s $z$-band DECam image (Prop. ID2016A-0190, PI Dey, observers Dustin Lang, Alistair Walker).
    \textbf{(d)} UT 2017 July 5 Las Cumbres Observatory Global Telescope (LCOGT) $3\times180$~s $R$-band archival image taken on the 2~m Faulkes Telescope North (LCOGT site code ``ogg''), Hawaii. Prop. ID FTPEPO2014A-004, user Americo Watkins. 
    \textbf{(e)} UT 2017 August 28 7$\times$120~s $r$-band images acquired with the 2~m Siding Spring Observatory (Australia) LCOGT telescope, site ``coj'' (Prop. ID FTPEPO2017AB-002, user ID Richard Mile). 
    \textbf{(f)} UT 2017 September 21 1$\times$100~s $g$-band + $1\times100$~s $r$-band 1.35~m SkyMapper telescope (Siding Spring). 
    \textbf{(g)} UT 2017 October 10 1$\times$100~s $r$-band SkyMapper image. 
    \textbf{(h)} UT 2024 April 3 1$\times$30~s $g$-band + 1$\times$30~s $r$-band ZTF images acquired with the 48'' Samuel Oschin Schmidt telescope (Palomar). 
    \textbf{(i)} UT 2024 April 30 2$\times$30~s $g$-band + 2$\times$30~s $z$-band ZTF exposures. 
    \textbf{(j)} UT 2024 May 14 1$\times$30~s $g$-band + 1$\times$30~s $r$-band ZTF exposures.}
    \end{figure}

In Figure \ref{fig:archival} we provide images of activity in archival data our team uncovered during our investigation. These data span both epochs discussed in this work.

\section{Orbital Parameters}

\setcounter{table}{1}

\begin{deluxetable*}{lrcc}
\label{tab:params}
\caption{\objname{} orbital parameters. Parameters and uncertainties (1-sigma) were acquired on UT 2024 February 28 from the JPL Horizons Small Body Database \citep{giorginiJPLOnLineSolar1996}.}
\tablehead{\colhead{Parameter} & \colhead{Value} & \colhead{Uncertainty} & \colhead{Units}}
\startdata
    Semi-major axis $a$ & 3.972 & 2.548$\times10^{-7}$ & au \\
    Eccentricity $e$ & 0.278 & 8.920$\times10^{-8}$ & - \\
    Inclination $i$ & 15.556 & 9.634$\times10^{-6}$ & deg \\
    Longitude of the ascending node $\Omega$ & 192.544 & 5.113$\times10^{-5}$ & deg \\
    Argument of perihelion $\omega$ & 53.537 & 8.722$\times10^{-5}$ & deg \\
    Mean anomaly $M$ & 321.244 & 5.835$\times10^{-5}$ & deg \\
    Perihelion distance $q$ & 2.865 & 4.141$\times10^{-7}$ & au \\
    Aphelion distance $Q$ & 5.080 & 3.258$\times10^{-7}$ & au \\
    Orbital period $P$ & 7.918 & 7.618$\times10^{-7}$ & yr \\
    UT Date of Perihelion & 2024 July 20 & - & -\\
    Tisserand parameter ($T_\mathrm{J}$) & 2.927 & - & -\\
    Minimum Orbit Intersection Distance of Jupiter (Jupiter MOID) & 0.354 & & au\\
\enddata
\end{deluxetable*}

In Table \ref{tab:params} we provide a listing of orbital parameters for \objname{}.


\bibliography{zotero}{}

\begin{thebibliography}{}
\expandafter\ifx\csname natexlab\endcsname\relax\def\natexlab#1{#1}\fi
\providecommand{\url}[1]{\href{#1}{#1}}
\providecommand{\dodoi}[1]{doi:~\href{http://doi.org/#1}{\nolinkurl{#1}}}
\providecommand{\doeprint}[1]{\href{http://ascl.net/#1}{\nolinkurl{http://ascl.net/#1}}}
\providecommand{\doarXiv}[1]{\href{https://arxiv.org/abs/#1}{\nolinkurl{https://arxiv.org/abs/#1}}}

\bibitem[{{Bellm} {et~al.}(2019{\natexlab{a}}){Bellm}, {Kulkarni}, {Graham}, {Dekany}, {Smith}, {Riddle}, {Masci}, {Helou}, {Prince}, {Adams}, {Barbarino}, {Barlow}, {Bauer}, {Beck}, {Belicki}, {Biswas}, {Blagorodnova}, {Bodewits}, {Bolin}, {Brinnel}, {Brooke}, {Bue}, {Bulla}, {Burruss}, {Cenko}, {Chang}, {Connolly}, {Coughlin}, {Cromer}, {Cunningham}, {De}, {Delacroix}, {Desai}, {Duev}, {Eadie}, {Farnham}, {Feeney}, {Feindt}, {Flynn}, {Franckowiak}, {Frederick}, {Fremling}, {Gal-Yam}, {Gezari}, {Giomi}, {Goldstein}, {Golkhou}, {Goobar}, {Groom}, {Hacopians}, {Hale}, {Henning}, {Ho}, {Hover}, {Howell}, {Hung}, {Huppenkothen}, {Imel}, {Ip}, {Ivezi{\'c}}, {Jackson}, {Jones}, {Juric}, {Kasliwal}, {Kaspi}, {Kaye}, {Kelley}, {Kowalski}, {Kramer}, {Kupfer}, {Landry}, {Laher}, {Lee}, {Lin}, {Lin}, {Lunnan}, {Giomi}, {Mahabal}, {Mao}, {Miller}, {Monkewitz}, {Murphy}, {Ngeow}, {Nordin}, {Nugent}, {Ofek}, {Patterson}, {Penprase}, {Porter}, {Rauch}, {Rebbapragada}, {Reiley}, {Rigault}, {Rodriguez}, {van Roestel},
  {Rusholme}, {van Santen}, {Schulze}, {Shupe}, {Singer}, {Soumagnac}, {Stein}, {Surace}, {Sollerman}, {Szkody}, {Taddia}, {Terek}, {Van Sistine}, {van Velzen}, {Vestrand}, {Walters}, {Ward}, {Ye}, {Yu}, {Yan}, \& {Zolkower}}]{ZTFCITATION}
{Bellm}, E.~C., {Kulkarni}, S.~R., {Graham}, M.~J., {et~al.} 2019{\natexlab{a}}, \pasp, 131, 018002, \dodoi{10.1088/1538-3873/aaecbe}

\bibitem[{{Bellm} {et~al.}(2019{\natexlab{b}}){Bellm}, {Kulkarni}, {Graham}, {Dekany}, {Smith}, {Riddle}, {Masci}, {Helou}, {Prince}, {Adams}, {Barbarino}, {Barlow}, {Bauer}, {Beck}, {Belicki}, {Biswas}, {Blagorodnova}, {Bodewits}, {Bolin}, {Brinnel}, {Brooke}, {Bue}, {Bulla}, {Burruss}, {Cenko}, {Chang}, {Connolly}, {Coughlin}, {Cromer}, {Cunningham}, {De}, {Delacroix}, {Desai}, {Duev}, {Eadie}, {Farnham}, {Feeney}, {Feindt}, {Flynn}, {Franckowiak}, {Frederick}, {Fremling}, {Gal-Yam}, {Gezari}, {Giomi}, {Goldstein}, {Golkhou}, {Goobar}, {Groom}, {Hacopians}, {Hale}, {Henning}, {Ho}, {Hover}, {Howell}, {Hung}, {Huppenkothen}, {Imel}, {Ip}, {Ivezi{\'c}}, {Jackson}, {Jones}, {Juric}, {Kasliwal}, {Kaspi}, {Kaye}, {Kelley}, {Kowalski}, {Kramer}, {Kupfer}, {Landry}, {Laher}, {Lee}, {Lin}, {Lin}, {Lunnan}, {Giomi}, {Mahabal}, {Mao}, {Miller}, {Monkewitz}, {Murphy}, {Ngeow}, {Nordin}, {Nugent}, {Ofek}, {Patterson}, {Penprase}, {Porter}, {Rauch}, {Rebbapragada}, {Reiley}, {Rigault}, {Rodriguez}, {van Roestel},
  {Rusholme}, {van Santen}, {Schulze}, {Shupe}, {Singer}, {Soumagnac}, {Stein}, {Surace}, {Sollerman}, {Szkody}, {Taddia}, {Terek}, {Van Sistine}, {van Velzen}, {Vestrand}, {Walters}, {Ward}, {Ye}, {Yu}, {Yan}, \& {Zolkower}}]{2019PASP..131a8002B}
---. 2019{\natexlab{b}}, \pasp, 131, 018002, \dodoi{10.1088/1538-3873/aaecbe}

\bibitem[{{Berthier} {et~al.}(2006){Berthier}, {Vachier}, {Thuillot}, {Fernique}, {Ochsenbein}, {Genova}, {Lainey}, \& {Arlot}}]{2006ASPC..351..367B}
{Berthier}, J., {Vachier}, F., {Thuillot}, W., {et~al.} 2006, in Astronomical Society of the Pacific Conference Series, Vol. 351, Astronomical Data Analysis Software and Systems XV, ed. C.~{Gabriel}, C.~{Arviset}, D.~{Ponz}, \& S.~{Enrique}, 367

\bibitem[{Berthier {et~al.}(2006)Berthier, Vachier, Thuillot, Fernique, Ochsenbein, Genova, Lainey, Arlot, Gabriel, Arviset, Ponz, \& Solano}]{berthierSkyBoTNewVO2006}
Berthier, J., Vachier, F., Thuillot, W., {et~al.} 2006, in Astronomical {{Data Analysis Software}} and {{Systems XV ASP Conference Series}}, Vol. 351 ({Orem, UT}: {Astronomical Society of the Pacific}), 367

\bibitem[{{Bowell} {et~al.}(1994){Bowell}, {Muinonen}, \& {Wasserman}}]{1994IAUS..160..477B}
{Bowell}, E., {Muinonen}, K., \& {Wasserman}, L.~H. 1994, in Asteroids, Comets, Meteors 1993, ed. A.~{Milani}, M.~{di Martino}, \& A.~{Cellino}, Vol. 160, 477--481

\bibitem[{Bradley {et~al.}(2023)Bradley, Sip{\H o}cz, Robitaille, Tollerud, Vin{\'{\i}}cius, Deil, Barbary, Wilson, Busko, Donath, G{\"u}nther, Cara, Lim, Me{\ss}linger, Conseil, Bostroem, Droettboom, Bray, Bratholm, Barentsen, Craig, Rathi, Pascual, Perren, Georgiev, de~Val-Borro, Kerzendorf, Bach, Quint, \& Souchereau}]{larry_bradley_2023_7946442}
Bradley, L., Sip{\H o}cz, B., Robitaille, T., {et~al.} 2023, astropy/photutils: 1.8.0, 1.8.0,  Zenodo, \dodoi{10.5281/zenodo.7946442}

\bibitem[{{Brown} {et~al.}(2013){Brown}, {Baliber}, {Bianco}, {Bowman}, {Burleson}, {Conway}, {Crellin}, {Depagne}, {De Vera}, {Dilday}, {Dragomir}, {Dubberley}, {Eastman}, {Elphick}, {Falarski}, {Foale}, {Ford}, {Fulton}, {Garza}, {Gomez}, {Graham}, {Greene}, {Haldeman}, {Hawkins}, {Haworth}, {Haynes}, {Hidas}, {Hjelstrom}, {Howell}, {Hygelund}, {Lister}, {Lobdill}, {Martinez}, {Mullins}, {Norbury}, {Parrent}, {Paulson}, {Petry}, {Pickles}, {Posner}, {Rosing}, {Ross}, {Sand}, {Saunders}, {Shobbrook}, {Shporer}, {Street}, {Thomas}, {Tsapras}, {Tufts}, {Valenti}, {Vander Horst}, {Walker}, {White}, \& {Willis}}]{2013PASP..125.1031B}
{Brown}, T.~M., {Baliber}, N., {Bianco}, F.~B., {et~al.} 2013, \pasp, 125, 1031, \dodoi{10.1086/673168}

\bibitem[{Chandler {et~al.}(2020)Chandler, Kueny, Trujillo, Trilling, \& Oldroyd}]{chandlerCometaryActivityDiscovered2020b}
Chandler, C.~O., Kueny, J.~K., Trujillo, C.~A., Trilling, D.~E., \& Oldroyd, W.~J. 2020, The Astrophysical Journal Letters, 892, L38, \dodoi{10/gg36xz}

\bibitem[{Chandler {et~al.}(2022)Chandler, Oldroyd, \& Trujillo}]{chandlerMigratoryOutburstingQuasiHilda2022}
Chandler, C.~O., Oldroyd, W.~J., \& Trujillo, C.~A. 2022, The Astrophysical Journal, 937, L2, \dodoi{10.3847/2041-8213/ac897a}

\bibitem[{Chandler {et~al.}(2024)Chandler, Trujillo, Oldroyd, Kueny, Burris, Hsieh, DeSpain, Sedaghat, Sheppard, Farrell, Trilling, Gustafsson, Magbanua, Mazzucato, Bosch, Shaw-Diaz, Gonano, Lamperti, da~Silva~Campos, Goodwin, Terentev, Dukes, \& Deen}]{Chandler_2024}
Chandler, C.~O., Trujillo, C.~A., Oldroyd, W.~J., {et~al.} 2024, The Astronomical Journal, 167, 156, \dodoi{10.3847/1538-3881/ad1de2}

\bibitem[{Correa-Otto {et~al.}(2023)Correa-Otto, García-Migani, \& Gil-Hutton}]{gil-hutton2023}
Correa-Otto, J., García-Migani, E., \& Gil-Hutton, R. 2023, Monthly Notices of the Royal Astronomical Society, 527, 876, \dodoi{10.1093/mnras/stad3234}

\bibitem[{da~Costa-Luis {et~al.}(2020)da~Costa-Luis, Larroque, Altendorf, Mary, Korobov, Yorav-Raphael, Ivanov, Bargull, Rodrigues, CHEN, Newey, James, Zugnoni, Pagel, mjstevens777, Dektyarev, Rothberg, Alexander, Panteleit, Dill, FichteFoll, HeoHeo, van Kemenade, McCracken, Nordlund, Nechaev, Desh, RedBug312, richardsheridan, \& Socialery}]{casper_da_costa_luis_2020_4293724}
da~Costa-Luis, C., Larroque, S.~K., Altendorf, K., {et~al.} 2020, {tqdm: A fast, Extensible Progress Bar for Python and CLI}, v4.54.0,  Zenodo, \dodoi{10.5281/zenodo.4293724}

\bibitem[{{Di Sisto} {et~al.}(2005){Di Sisto}, {Brunini}, {Dirani}, \& {Orellana}}]{2005Icar..174...81D}
{Di Sisto}, R.~P., {Brunini}, A., {Dirani}, L.~D., \& {Orellana}, R.~B. 2005, \icarus, 174, 81, \dodoi{10.1016/j.icarus.2004.10.024}

\bibitem[{{Di Sisto} {et~al.}(2019){Di Sisto}, {Ramos}, \& {Gallardo}}]{2019Icar..319..828D}
{Di Sisto}, R.~P., {Ramos}, X.~S., \& {Gallardo}, T. 2019, \icarus, 319, 828, \dodoi{10.1016/j.icarus.2018.10.029}

\bibitem[{{Flaugher} {et~al.}(2015){Flaugher}, {Diehl}, {Honscheid}, {Abbott}, {Alvarez}, {Angstadt}, {Annis}, {Antonik}, {Ballester}, {Beaufore}, {Bernstein}, {Bernstein}, {Bigelow}, {Bonati}, {Boprie}, {Brooks}, {Buckley-Geer}, {Campa}, {Cardiel-Sas}, {Castander}, {Castilla}, {Cease}, {Cela-Ruiz}, {Chappa}, {Chi}, {Cooper}, {da Costa}, {Dede}, {Derylo}, {DePoy}, {de Vicente}, {Doel}, {Drlica-Wagner}, {Eiting}, {Elliott}, {Emes}, {Estrada}, {Fausti Neto}, {Finley}, {Flores}, {Frieman}, {Gerdes}, {Gladders}, {Gregory}, {Gutierrez}, {Hao}, {Holland}, {Holm}, {Huffman}, {Jackson}, {James}, {Jonas}, {Karcher}, {Karliner}, {Kent}, {Kessler}, {Kozlovsky}, {Kron}, {Kubik}, {Kuehn}, {Kuhlmann}, {Kuk}, {Lahav}, {Lathrop}, {Lee}, {Levi}, {Lewis}, {Li}, {Mandrichenko}, {Marshall}, {Martinez}, {Merritt}, {Miquel}, {Mu{\~n}oz}, {Neilsen}, {Nichol}, {Nord}, {Ogando}, {Olsen}, {Palaio}, {Patton}, {Peoples}, {Plazas}, {Rauch}, {Reil}, {Rheault}, {Roe}, {Rogers}, {Roodman}, {Sanchez}, {Scarpine}, {Schindler}, {Schmidt},
  {Schmitt}, {Schubnell}, {Schultz}, {Schurter}, {Scott}, {Serrano}, {Shaw}, {Smith}, {Soares-Santos}, {Stefanik}, {Stuermer}, {Suchyta}, {Sypniewski}, {Tarle}, {Thaler}, {Tighe}, {Tran}, {Tucker}, {Walker}, {Wang}, {Watson}, {Weaverdyck}, {Wester}, {Woods}, {Yanny}, \& {DES Collaboration}}]{DecamCite}
{Flaugher}, B., {Diehl}, H.~T., {Honscheid}, K., {et~al.} 2015, \aj, 150, 150, \dodoi{10.1088/0004-6256/150/5/150}

\bibitem[{Fraser {et~al.}(2022)Fraser, Dones, Volk, Womack, \& Nesvorný}]{fraser2022transition}
Fraser, W.~C., Dones, L., Volk, K., Womack, M., \& Nesvorný, D. 2022, The Transition from the Kuiper Belt to the Jupiter-Family (Comets).
\newblock \doarXiv{2210.16354}

\bibitem[{García-Migani \& Gil-Hutton(2018)}]{GARCIAMIGANI201812}
García-Migani, E., \& Gil-Hutton, R. 2018, Planetary and Space Science, 160, 12, \dodoi{https://doi.org/10.1016/j.pss.2018.03.011}

\bibitem[{{Gil-Hutton} \& {Garc{\'\i}a-Migani}(2016)}]{2016A&A...590A.111G}
{Gil-Hutton}, R., \& {Garc{\'\i}a-Migani}, E. 2016, \aap, 590, A111, \dodoi{10.1051/0004-6361/201628184}

\bibitem[{Giorgini {et~al.}(1996)Giorgini, Yeomans, Chamberlin, Chodas, Jacobson, Keesey, Lieske, Ostro, Standish, \& Wimberly}]{giorginiJPLOnLineSolar1996}
Giorgini, J.~D., Yeomans, D.~K., Chamberlin, A.~B., {et~al.} 1996, American Astronomical Society, 28, 25.04

\bibitem[{Gwyn {et~al.}(2012)Gwyn, Hill, \& Kavelaars}]{gwyn2012ssos}
Gwyn, S.~D., Hill, N., \& Kavelaars, J. 2012, Publications of the Astronomical Society of the Pacific, 124, 579

\bibitem[{Harris {et~al.}(2020)Harris, Millman, {van der Walt}, Gommers, Virtanen, Cournapeau, Wieser, Taylor, Berg, Smith, Kern, Picus, Hoyer, {van Kerkwijk}, Brett, Haldane, {del R{\'i}o}, Wiebe, Peterson, {G{\'e}rard-Marchant}, Sheppard, Reddy, Weckesser, Abbasi, Gohlke, \& Oliphant}]{harrisArrayProgrammingNumPy2020}
Harris, C.~R., Millman, K.~J., {van der Walt}, S.~J., {et~al.} 2020, Nature, 585, 357, \dodoi{10.1038/s41586-020-2649-2}

\bibitem[{Hsieh {et~al.}(2015{\natexlab{a}})Hsieh, Denneau, Wainscoat, Sch{\"o}rghofer, Bolin, Fitzsimmons, Jedicke, Kleyna, Micheli, Vere{\v s}, Kaiser, Chambers, Burgett, Flewelling, Hodapp, Magnier, Morgan, Price, Tonry, \& Waters}]{hsiehMainbeltCometsPanSTARRS12015}
Hsieh, H.~H., Denneau, L., Wainscoat, R.~J., {et~al.} 2015{\natexlab{a}}, Icarus, 248, 289, \dodoi{10.1016/j.icarus.2014.10.031}

\bibitem[{Hsieh {et~al.}(2015{\natexlab{b}})Hsieh, Denneau, Wainscoat, Schörghofer, Bolin, Fitzsimmons, Jedicke, Kleyna, Micheli, Vereš, Kaiser, Chambers, Burgett, Flewelling, Hodapp, Magnier, Morgan, Price, Tonry, \& Waters}]{HSIEH2015289}
---. 2015{\natexlab{b}}, Icarus, 248, 289, \dodoi{https://doi.org/10.1016/j.icarus.2014.10.031}

\bibitem[{{Huehnerhoff} {et~al.}(2016){Huehnerhoff}, {Ketzeback}, {Bradley}, {Dembicky}, {Doughty}, {Hawley}, {Johnson}, {Klaene}, {Leon}, {McMillan}, {Owen}, {Sayres}, {Sheen}, \& {Shugart}}]{2016SPIE.9908E..5HH}
{Huehnerhoff}, J., {Ketzeback}, W., {Bradley}, A., {et~al.} 2016, in Society of Photo-Optical Instrumentation Engineers (SPIE) Conference Series, Vol. 9908, Ground-based and Airborne Instrumentation for Astronomy VI, ed. C.~J. {Evans}, L.~{Simard}, \& H.~{Takami}, 99085H, \dodoi{10.1117/12.2234214}

\bibitem[{Hunter(2007)}]{hunterMatplotlib2DGraphics2007}
Hunter, J.~D. 2007, Computing in Science \& Engineering, 9, 90, \dodoi{10.1109/MCSE.2007.55}

\bibitem[{Jewitt {et~al.}(2015)Jewitt, Hsieh, \& Agarwal}]{jewittActiveAsteroids2015}
Jewitt, D., Hsieh, H., \& Agarwal, J. 2015, in Asteroids {{IV}} ({Tucson, Arizona}: {University of Arizona Press}), 221--241

\bibitem[{Jewitt \& Hsieh(2022)}]{jewitt2022asteroidcomet}
Jewitt, D., \& Hsieh, H.~H. 2022, The Asteroid-Comet Continuum.
\newblock \doarXiv{2203.01397}

\bibitem[{Joye(2006)}]{joyeNewFeaturesSAOImage2006}
Joye, W.~A. 2006, in Astronomical {{Data Analysis Software}} and {{Systems XV ASP Conference Series}}, Vol. 351, 574--

\bibitem[{Keller {et~al.}(2007)Keller, Schmidt, Bessell, Conroy, Francis, Granlund, Kowald, Oates, Martin-Jones, Preston, \& et~al.}]{Keller2007}
Keller, S.~C., Schmidt, B.~P., Bessell, M.~S., {et~al.} 2007, Publications of the Astronomical Society of Australia, 24, 1–12, \dodoi{10.1071/AS07001}

\bibitem[{Lang {et~al.}(2010)Lang, Hogg, Mierle, Blanton, \& Roweis}]{langAstrometryNetBlind2010}
Lang, D., Hogg, D.~W., Mierle, K., Blanton, M., \& Roweis, S. 2010, Astronomical Journal, 139, 1782, \dodoi{10.1088/0004-6256/139/5/1782}

\bibitem[{{Levine} {et~al.}(2022){Levine}, {Ellsworth-Bowers}, {Bida}, {Hamilton}, \& {Kuehn}}]{2022SPIE12182E..27L}
{Levine}, S.~E., {Ellsworth-Bowers}, T., {Bida}, T.~A., {Hamilton}, R., \& {Kuehn}, K. 2022, in Society of Photo-Optical Instrumentation Engineers (SPIE) Conference Series, Vol. 12182, Ground-based and Airborne Telescopes IX, ed. H.~K. {Marshall}, J.~{Spyromilio}, \& T.~{Usuda}, 1218227, \dodoi{10.1117/12.2629797}

\bibitem[{{Levison}(1996)}]{1996ASPC..107..173L}
{Levison}, H.~F. 1996, in Astronomical Society of the Pacific Conference Series, Vol. 107, Completing the Inventory of the Solar System, ed. T.~{Rettig} \& J.~M. {Hahn}, 173--191

\bibitem[{Lilly {et~al.}(2023)Lilly, Jevčák, Schambeau, Volk, Steckloff, Hsieh, Fernandez, Bauer, Weryk, \& Wainscoat}]{lilly2023semimajor}
Lilly, E., Jevčák, P., Schambeau, C., {et~al.} 2023, Semi-major Axis Jumps as the Activity Trigger in Centaurs and High-Perihelion Jupiter Family Comets.
\newblock \doarXiv{2312.06847}

\bibitem[{Moskovitz {et~al.}(2021)Moskovitz, Burt, Schottland, Wasserman, Bailen, Grimm, \& Granvik}]{moskovitzAstorbDatabaseLowell2021}
Moskovitz, N., Burt, B., Schottland, R., {et~al.} 2021, AAS Division of Planetary Science meeting \#53, id. 101.04, 53, 101.04

\bibitem[{Ochsenbein {et~al.}(2000)Ochsenbein, Bauer, \& Marcout}]{ochsenbeinVizieRDatabaseAstronomical2000}
Ochsenbein, F., Bauer, P., \& Marcout, J. 2000, Astronomy and Astrophysics Supplement, 143, 23, \dodoi{10/fb95hg}

\bibitem[{{Oldroyd} {et~al.}(2023){Oldroyd}, {Chandler}, {Trujillo}, {Sheppard}, {Hsieh}, {Kueny}, {Burris}, {DeSpain}, {Farrell}, {Mazzucato}, {Bosch}, {Shaw-Diaz}, \& {Gonano}}]{2023ApJ...957L...1O}
{Oldroyd}, W.~J., {Chandler}, C.~O., {Trujillo}, C.~A., {et~al.} 2023, \apjl, 957, L1, \dodoi{10.3847/2041-8213/acfcbc}

\bibitem[{Reback {et~al.}(2022)Reback, {jbrockmendel}, McKinney, den Bossche, Augspurger, Roeschke, Hawkins, Cloud, {gfyoung}, Sinhrks, Hoefler, Klein, Petersen, Tratner, She, Ayd, Naveh, Darbyshire, Garcia, Shadrach, Schendel, Hayden, Saxton, Gorelli, Li, Zeitlin, Jancauskas, McMaster, W{\"o}rtwein, \& Battiston}]{rebackPandasdevPandasPandas2022}
Reback, J., {jbrockmendel}, McKinney, W., {et~al.} 2022, Pandas-Dev/Pandas: {{Pandas}} 1.4.2, Zenodo, \dodoi{10.5281/zenodo.6408044}

\bibitem[{{Rein} \& {Liu}(2012)}]{2012A&A...537A.128R}
{Rein}, H., \& {Liu}, S.~F. 2012, \aap, 537, A128, \dodoi{10.1051/0004-6361/201118085}

\bibitem[{{Rein} \& {Spiegel}(2015)}]{2015MNRAS.446.1424R}
{Rein}, H., \& {Spiegel}, D.~S. 2015, \mnras, 446, 1424, \dodoi{10.1093/mnras/stu2164}

\bibitem[{Robitaille {et~al.}(2013)Robitaille, Tollerud, Greenfield, Droettboom, Bray, Aldcroft, Davis, Ginsburg, {Price-Whelan}, Kerzendorf, Conley, Crighton, Barbary, Muna, Ferguson, Grollier, Parikh, Nair, G{\"u}nther, Deil, Woillez, Conseil, Kramer, Turner, Singer, Fox, Weaver, Zabalza, Edwards, Azalee~Bostroem, Burke, Casey, Crawford, Dencheva, Ely, Jenness, Labrie, Lim, Pierfederici, Pontzen, Ptak, Refsdal, Servillat, \& Streicher}]{robitailleAstropyCommunityPython2013}
Robitaille, T.~P., Tollerud, E.~J., Greenfield, P., {et~al.} 2013, Astronomy \& Astrophysics, 558, A33, \dodoi{10/gfvntd}

\bibitem[{{Toth}(2006)}]{2006A&A...448.1191T}
{Toth}, I. 2006, \aap, 448, 1191, \dodoi{10.1051/0004-6361:20053492}

\bibitem[{Virtanen {et~al.}(2020)Virtanen, Gommers, Oliphant, Haberland, Reddy, Cournapeau, Burovski, Peterson, Weckesser, Bright, {van der Walt}, Brett, Wilson, Millman, Mayorov, Nelson, Jones, Kern, Larson, Carey, Polat, Feng, Moore, VanderPlas, Laxalde, Perktold, Cimrman, Henriksen, Quintero, Harris, Archibald, Ribeiro, Pedregosa, \& {van Mulbregt}}]{virtanenSciPyFundamentalAlgorithms2020}
Virtanen, P., Gommers, R., Oliphant, T.~E., {et~al.} 2020, Nature Methods, 17, 261, \dodoi{10.1038/s41592-019-0686-2}

\bibitem[{{Weisenburger} {et~al.}(2017){Weisenburger}, {Huehnerhoff}, {Levesque}, \& {Massey}}]{2017JOSS....2..102L}
{Weisenburger}, K., {Huehnerhoff}, J., {Levesque}, E., \& {Massey}, P. 2017, The Journal of Open Source Software, 2, 102, \dodoi{10.21105/joss.00102}

\bibitem[{Weisenburger {et~al.}(2017)Weisenburger, Huehnerhoff, Levesque, \& Massey}]{weisenburger2017acronym}
Weisenburger, K.~L., Huehnerhoff, J., Levesque, E.~M., \& Massey, P. 2017, J. Open Source Softw., 2, 102

\bibitem[{{West} {et~al.}(1997){West}, {Nagel}, {Harvey}, {Brar}, {Phillips}, {Ray}, {Trebisky}, {Cromwell}, {Woolf}, {Corbally}, {Boyle}, {Blanco}, \& {Otten}}]{1997SPIE.2871...74W}
{West}, S.~C., {Nagel}, R.~H., {Harvey}, D.~A., {et~al.} 1997, in Society of Photo-Optical Instrumentation Engineers (SPIE) Conference Series, Vol. 2871, Optical Telescopes of Today and Tomorrow, ed. A.~L. {Ardeberg}, 74--85, \dodoi{10.1117/12.268968}

\end{thebibliography}
\bibliographystyle{aasjournal}



\end{document}